\documentclass[amsmath,amsfonts,aps,twocolumn,twoside,superscriptaddress]{revtex4-2}
\pdfoutput=1

\usepackage[utf8]{inputenc}

\usepackage{amssymb}
\usepackage{amsfonts}
\usepackage{amsmath}
\usepackage{graphicx}
\usepackage{indentfirst}
\usepackage{dsfont}

\allowdisplaybreaks[4]

\usepackage{mathtools}

\usepackage{enumerate}
\usepackage{url}

\usepackage{array}
\usepackage{multirow}

\usepackage{amsthm}

\usepackage{tikz,pgfplots}
\definecolor{darkblue}{rgb}{0.15,0.35,0.55}
\definecolor{reddish}{rgb}{.8, 0.2, 0.2}
\definecolor{plotblue}{RGB}{0,119,187}
\definecolor{plotgreen}{RGB}{0,153,136}
\definecolor{plotorange}{RGB}{238,119,51}
\definecolor{plotmagenta}{RGB}{238,51,119}
\definecolor{plotgray}{RGB}{128,128,128}
\definecolor{plotcyan}{RGB}{51,187,238}
\definecolor{plotred}{RGB}{204,51,17}
\usepgfplotslibrary{fillbetween}

\long\def\ca#1\cb{} 

\newcommand{\becs}{\begin{cases}}
	\newcommand{\bem}{\begin{matrix}}

		\newcommand{\dya}[1]{|#1\rangle\langle#1|}
		\newcommand{\dyad}[2]{|#1\rangle\langle#2|}
		
		\newcommand{\encs}{\end{cases}}
	\newcommand{\enm}{\end{matrix}}

\newcommand{\ket}[1]{|#1\rangle }

\newcommand{\ot}{\otimes }

\newcommand{\Tr}{{\rm Tr}}

\newcommand{\AC}{{\mathcal A}}
\newcommand{\BC}{{\mathcal B}}
\newcommand{\CC}{{\mathcal C}}
\newcommand{\DC}{{\mathcal D}}
\newcommand{\EC}{{\mathcal E}}
\newcommand{\FC}{{\mathcal F}}

\newcommand{\HC}{{\mathcal H}}
\newcommand{\IC}{{\mathcal I}}

\newcommand{\KC}{{\mathcal K}}

\newcommand{\MC}{{\mathcal M}}
\newcommand{\NC}{{\mathcal N}}

\newcommand{\PC}{{\mathcal P}}
\newcommand{\QC}{{\mathcal Q}}
\newcommand{\RC}{{\mathcal R}}
\newcommand{\SC}{{\mathcal S}}

\newcommand{\gm}{\gamma }

\newcommand{\dl}{\delta }
\newcommand{\Dl}{\Delta }
\newcommand{\ep}{\epsilon}

\newcommand{\lm}{\lambda }

\newcommand{\sg}{\sigma }

\DeclareMathOperator{\tr}{tr}
\newcommand{\one}{\mathds{1}}

\DeclareMathOperator{\id}{id}

\usepackage[pdfpagelabels, linktocpage=true,pdfusetitle]{hyperref}
\hypersetup{
	colorlinks=true,
	citecolor=darkblue,
	linkcolor=reddish,
	urlcolor=darkblue,
	pdfauthor={},
	pdfsubject={}
}

\usepackage{cleveref}

\usepackage[normalem]{ulem}

\newcommand{\aux}{\KC}

\begin{document}
\title{Generic nonadditivity of quantum capacity in simple channels}

\author{Felix Leditzky}
\email{leditzky@illinois.edu}
\affiliation{Department of Mathematics and IQUIST, University of Illinois at Urbana-Champaign, Urbana, IL 61801, USA}
\affiliation{Institute for Quantum Computing, and Department of Combinatorics \& Optimization, University of Waterloo, Waterloo, ON N2L 3G1, Canada}
\affiliation{Perimeter Institute for Theoretical Physics, Waterloo, ON N2L 2Y5, Canada}

\author{Debbie Leung}
\email{wcleung@uwaterloo.ca}
\affiliation{Institute for Quantum Computing, and Department of Combinatorics \& Optimization, University of Waterloo, Waterloo, ON N2L 3G1, Canada}
\affiliation{Perimeter Institute for Theoretical Physics, Waterloo, ON N2L 2Y5, Canada}

\author{Vikesh Siddhu}
\email{vsiddhu@protonmail.com}
\altaffiliation{Present Address: IBM Quantum}
\affiliation{JILA, University of Colorado/NIST, 440 UCB, Boulder, CO 80309, USA}
\affiliation{Department of Physics and Quantum Computing Group, Carnegie Mellon University, Pittsburgh, PA, 15213, USA}

\author{Graeme Smith}
\email{graeme.smith@colorado.edu}
\affiliation{JILA, University of Colorado/NIST, 440 UCB, Boulder, CO 80309, USA}
\affiliation{Department of Physics and Center for Theory of Quantum Matter, University of Colorado, Boulder, Colorado 80309, USA}

\author{John A. Smolin}
\email{smolin@us.ibm.com}
\affiliation{IBM Quantum, IBM T.J.~Watson Research Center, Yorktown Heights, New York 10598, USA}

\begin{abstract}    
    Determining capacities of quantum channels is a fundamental question in
    quantum information theory.  Despite having rigorous coding theorems
    quantifying the flow of information across quantum channels, their
    capacities are poorly understood due to super-additivity effects.  Studying
    these phenomena is important for deepening our understanding of quantum
    information, yet simple and clean examples of super-additive channels are
    scarce.  
    Here we study a family of channels called platypus channels. Its
    simplest member, a qutrit channel, is shown to display super-additivity of
    coherent information when used jointly with a variety of qubit channels.
    Higher-dimensional family members display super-additivity of
    quantum capacity together with an erasure channel.
    Subject to the ``spin-alignment conjecture'' introduced in the companion
    paper \cite{paper1IEEE}, our results on super-additivity of quantum capacity
    extend to lower-dimensional channels as well as larger parameter ranges.
    In particular, super-additivity occurs between two weakly additive channels
    each with large capacity on their own, in stark contrast to previous
    results.  Remarkably, a single, novel transmission strategy achieves
    super-additivity in all examples.  Our results show that super-additivity
    is much more prevalent than previously thought.  It can occur across a wide
    variety of channels, even when both participating channels have large
    quantum capacity.
\end{abstract}
\maketitle

\paragraph*{Introduction.}

A central aim of quantum information theory is to find out how much information
a noisy quantum channel can transmit reliably---to find a quantum channel’s
capacity~\cite{BennettShor04, Holevo20}.  In fact, a quantum channel has many
capacities, depending on what sorts of information are to be transmitted and
what additional resources are on hand.  The primary capacities of a quantum
channel are the classical~\cite{Holevo73a, Holevo98, SchumacherWestmoreland97},
private~\cite{BennettBrassard84, CaiWinterEA04, Devetak05}, and quantum
capacities~\cite{BennettDiVincenzoEA96, Lloyd97, BarnumNielsenEA98,
BarnumKnillEA00, Shor02a, Devetak05}.  This paper focuses on \emph{unassisted}
capacities, when no
additional resources (such as free entanglement) are available.  

The theory of quantum capacities is far richer and more complex than the
corresponding classical theory~\cite{Shannon48, CoverThomas01}. This richness
includes many synergies and surprises: super-additivity of coherent
information~\cite{ShorJohn96, DiVincenzoShorEA98, FernWhaley08, SmithSmolin07,
CubittElkoussEA15, LimLee18, LeditzkyLeungEA18, BauschLeditzky19,
BauschLeditzky20, SiddhuGriffiths20, Siddhu20, Siddhu21, Filippov21,
SidhardhAlimuddinEA22, KoudiaCacciapuotiEA21}, private
information~\cite{SmithSmolin09a, SmithSmolin09, ElkoussStrelchuk15}, and
Holevo information~\cite{Hastings09}, superactivation of quantum
capacity~\cite{SmithYard08, Oppenheim08, SmithSmolinEA11, BrandaoOppenheimEA12,
LimTakagiEA19}, and private communication at a rate above the quantum
capacity~\cite{HorodeckiHorodeckiEA05,LeungLiEA14}.  Over the past two decades,
there have been numerous exciting discoveries about these phenomena, but they
remain mysterious.  As a result, we don't have a theory of how to best
communicate with quantum channels, and can't answer many of the sorts of
questions classical information theory does. For example, in quantum
information theory random codes can be suboptimal, and we can only evaluate
capacities in special cases~\cite{BennettDiVincenzoEA97,Shor02,King02,King03,DevetakShor05, King06,KingMatsumotoEA07,Smith08,Watanabe12}.  Our
understanding of error correction in the quantum setting is thus incomplete,
whether the data is classical, private, or quantum.

Any quantum channel $\BC$ can be expressed as an isometry  $J\colon A \mapsto
BE$ followed by a partial trace over the environment $E$: $\BC(\rho) = \Tr_E (J
\rho J^{\dag})$.
Physically, it means that quantum noise arises from sharing the unclonable
quantum data with the environment which is subsequently lost (i.e., traced out).  
Therefore, to understand quantum transmission we must also consider the
environment's view of the channel, known as the complementary channel:
$\BC^c(\rho) = \Tr_B (J \rho J^{\dag})$. Together, the channel and its
complement allow us to define the coherent information of a channel $\BC$ on an
input state $\rho$ as $\Dl(\BC,\rho):= S\big( \BC (\rho) \big) - S \big( \BC^c
(\rho) \big)$, where $S(\sigma) = -\tr (\sigma\log \sigma)$ is the von Neumann
entropy of $\sigma$.
Mathematically, the coherent information signifies how much more information
about the input is available in system $B$ than in system $E$.  Operationally,
a random coding argument shows that indeed, for any input state $\rho$, the
quantity $\Dl(\BC,\rho)$ is an achievable rate for quantum
transmission~\cite{BarnumNielsenEA98,BarnumKnillEA00,Lloyd97,Shor02a,Devetak05}.
Maximizing over all inputs $\rho$ gives the channel coherent information
$\QC^{(1)}(\BC)$.

If the channel coherent information is additive, that is, $\QC^{(1)}(\BC_1
\otimes \BC_2) = \QC^{(1)}(\BC_1) + \QC^{(1)}(\BC_2)$ for any two channels
$\BC_1$ and $\BC_2$, then the theory of quantum capacity will resemble its
classical analogue.  
However, a rich theory of quantum capacity originates from two distinct notions
of nonadditivity: violations of \emph{weak additivity} and violations of
\emph{strong additivity}.

We first discuss violations of \emph{weak additivity}.  
The quantum capacity can be expressed as
\cite{SchumacherNielsen96,BarnumNielsenEA98,BarnumKnillEA00,Lloyd97,Shor02a,Devetak05}
\begin{align}
    \QC(\BC) = \lim_{n\rightarrow \infty} \frac{1}{n} \QC^{(1)}(\BC^{\ot n}),
    \label{eq:chanCap-main}
\end{align}
where $\BC^{\ot n}$ is the $n$-fold tensor product of $\BC$.
If $\QC^{(1)}(\BC^{\ot n}) = n \QC^{(1)}(\BC)$ for all $n\in\mathbb{N}$, we
say that $\BC$ has weakly additive coherent information, in which case
$\QC(\BC) = \QC^{(1)}(\BC)$.
However, there are channels $\BC$ for which
$\QC^{(1)}(\BC^{\ot n}) > n \QC^{(1)}(\BC)$ holds for some $n$
\cite{ShorJohn96, DiVincenzoShorEA98, FernWhaley08, SmithSmolin07,
CubittElkoussEA15, LeditzkyLeungEA18, BauschLeditzky19, BauschLeditzky20,
SiddhuGriffiths20, Siddhu21, Filippov21, SidhardhAlimuddinEA22}.  
Thus, the $n\rightarrow \infty$ limit is in general required in the
above \emph{regularized} expression for the quantum capacity.
When a channel does \emph{not} have weakly additive coherent
information, special quantum codes can outperform the
classical-inspired random coding strategy achieved by $\QC^{(1)}$.  
This
unbounded optimization also means that we can rarely determine the quantum capacity
of a quantum channel.

The second notion of nonadditivity, violations of strong additivity, can be phrased as follows.
For two channels $\BC_1$ and $\BC_2$, we have the general inequality
\begin{align}
    \QC^{(1)}(\BC_1 \otimes \BC_2) \geq \QC^{(1)}(\BC_1) + \QC^{(1)}(\BC_2) \,.\label{eq:CI-ineq}
\end{align}
Letting $\BC_1$ be a fixed channel, if equality in \eqref{eq:CI-ineq} holds for
all channels $\BC_2$, we say that $\BC_1$ has strongly additive coherent
information. In this case, the quantum capacity satisfies $\QC(\BC_1 \otimes
\BC_2)=\QC(\BC_1)+\QC(\BC_2)$.
Note that strong additivity implies weak additivity. 
Violations of strong additivity imply that two different channels can have
strictly superadditive coherent information, or even capacity.
As a result, not only do we not know the capacity of most quantum channels, we also do not
know when two channels used jointly can have capacity exceeding
the sum of the individual channels.  
A more general notion of a channel's capability to transmit quantum data thus
depends on the details of other resources available~\cite{SmithSmolinEA08,SmithYard08,YangWinter16}, and does not necessarily coincide with its capacity, a drastic
deviation from the classical theory.

Similar to the quantum capacity, a channel's private and classical
capacities can be defined as the highest rates of faithful
transmission of private and classical information, respectively;
expressions analogous to~\eqref{eq:chanCap-main} are
known~\cite{Devetak05,CaiWinterEA04,Holevo98,SchumacherWestmoreland97}.
Both capacities require regularized expressions
\cite{SmithRenesEA08,Hastings09}, and the private capacity can be shown to be
non-additive for some channels~\cite{LiWinterEA09, SmithSmolin09a}.  

For classical capacity, the underlying information quantity is the Holevo
information, which was conjectured to be additive for a long time.  In fact,
strong additivity was proved for certain channels such as entanglement-breaking
\cite{Shor02}, depolarizing \cite{King03}, Hadamard
\cite{King06,KingMatsumotoEA07}, and unital qubit channels \cite{King02}.  As a
result, for these channels the classical capacity completely characterizes
their ability to faithfully send classical information.
Furthermore, the only known proofs of violation of weak additivity of the
Holevo information \cite{Hastings09,fukuda2010comments,brandao2010hastings} are
based on random channel constructions and no explicit example has been found
yet~\cite{Hastings09, Holevo20}.  It is still open if the classical capacity
can be non-additive.  
It is furthermore unclear if additivity is more prevalent for classical data
transmission, or if proofs are simply harder to come by since the
Holevo information involves a more complex optimization compared to
coherent information.

The situation for quantum information transmission is quite different.  There is
a plethora of concrete channels with super-additive coherent information
~\cite{ShorJohn96, DiVincenzoShorEA98, FernWhaley08, SmithSmolin07,
LeditzkyLeungEA18, BauschLeditzky19, BauschLeditzky20, SiddhuGriffiths20,
Siddhu20, Siddhu21, Filippov21, SidhardhAlimuddinEA22}.
The only known class of channels with strongly additive coherent
information are the entanglement-breaking channels, but they
are somewhat trivial -- their quantum capacity is zero.  
Degradable channels \cite{DevetakShor05, CubittRuskaiEA08} have weakly additive
coherent information, and two degradable channels have additive
coherent information, yet surprisingly degradability does not imply
strong additivity for a channel.
Even weakly additive channels like some
\mbox{(anti-)}degradable \cite{DevetakShor05} and PPT channels
\cite{HorodeckiHorodeckiEA00} may have super-additive quantum capacity in
combination with suitable
channels~\cite{SmithYard08,LiWinterEA09,BrandaoOppenheimEA12}.  A common
feature in these violations of strong additivity is that one or both of the
channels are manifestly noisy, that is, with vanishing or small quantum
capacity.  Most of these proofs come from a qualitative inability for
the channels to transmit quantum data; in addition, nearly noiseless
channels are indeed limited in their non-additivity \cite{LeditzkyLeungEA18a}.

In this paper, we provide qualitatively new examples of super-additivity of
quantum capacity.  The phenomenon seems prevalent, does not involve
channels engineered to exhibit the effect, and can involve pairs of
channels with large quantum capacity.
Our findings show an even more complex landscape of non-additivity than
hitherto appreciated.  
Yet, our channels and the proofs are simple, and thus we hope they improve
our understanding of the subject.

\paragraph*{Main results.}

Our first main result is that a simple qutrit `platypus channel',
defined via eq.~\eqref{eq:Ns-iso} below,
violates strong additivity of coherent information when used together with a
variety of simple and well-known qubit channels such as the erasure, amplitude
damping, depolarizing, and even randomly constructed qubit channels.
Even more remarkably, the same simple code achieves non-additivity in all
cases.  Our findings strongly suggest that super-additivity is much more
prevalent and generic than previously thought.    

Second, as proved in our companion paper~\cite{paper1IEEE}, platypus channels have weakly additive coherent information if the spin alignment conjecture
introduced in \cite{paper1IEEE} holds.  As the erasure channel and the amplitude
damping channel also have weakly additive coherent information, we have an
example of non-additivity of quantum capacity between two weakly additive
channels.  The only known prior example revolves around
superactivation~\cite{SmithYard08}, and requires substantial fine-tuning to
demonstrate the effect.  In contrast, our channel requires no such tuning, and
both channels exhibit non-additivity over a wide range of parameters, including
regimes where both channels have substantial capacity themselves.

Third, we show that higher-dimensional platypus channels have similar
non-additive behavior.  In particular, when used jointly with a
higher-dimensional erasure channel, it exhibits super-additivity of quantum
capacity unconditionally, i.e., without relying on the spin alignment
conjecture.  The underlying mechanism at work achieving all of these
non-additivity results is qualitatively different from previous results
in~\cite{SmithYard08,LiWinterEA09,BrandaoOppenheimEA12}, as explained in the
Discussion section.

In the following paragraphs we discuss our main results; see the
\hyperref[supp]{Supplementary information}
for additional
details.  MATLAB and Python code used to obtain the numerical results mentioned
above will be made available at \cite{github}.

\paragraph*{The qutrit platypus channel.}
The qutrit `platypus channel' $\NC_s$ is defined by the following
isometry $F_s: \HC_a \mapsto \HC_b \ot \HC_c$:
\begin{align}
\begin{aligned}
	    F_s \ket{0} &= \sqrt{s} \ket{0} \ot \ket{0} + \sqrt{1-s} \ket{1}\ot \ket{1} \\
    F_s \ket{1} &= \ket{2} \ot \ket{0}\\
    F_s \ket{2} &= \ket{2} \ot \ket{1},
\end{aligned}
\label{eq:Ns-iso}
\end{align}
where $0 \leq s \leq 1/2$, and the input $\HC_a$, output $\HC_b$ and
environment $\HC_c$ have dimension $3,3$, and $2$, respectively.
This channel~\cite{Siddhu21, WangDuan18} is extensively studied in the
companion paper~\cite{paper1IEEE}.  From~\cite{Siddhu21, paper1IEEE}, the channel
coherent information is always positive and can be attained on inputs of the
form $\sg(u) \coloneqq(1-u) \dya{0} + u \dya{2}$: $$\QC^{(1)}(\NC_s) =
\max_{u\in[0,1]} \Dl(\NC_s,\sg(u))>0.$$ Conditioned on the spin-alignment
conjecture~(SAC) formulated in \cite{paper1IEEE}, the channel coherent information
$\QC^{(1)}(\NC_s)$ can be proved to be weakly additive, and thus $\QC(\NC_s) =
\QC^{(1)}(\NC_s)$.  Without the SAC, we have the upper bound $\QC(\NC_s) \leq
\log (1 + \sqrt{1-s})$.

\begin{figure}[t]
	\begin{tikzpicture}
	\begin{axis}[
	xlabel=$s$,
	xmin = 0,
	xmax = 0.5,
	ymin = 0,
	ymax = 0.08,
	scale=1.1,
	every axis plot/.append style={line width=1.5pt},
	legend cell align={left},
	legend style={at = {(0.025,0.975)},anchor = north west,/tikz/every even column/.append style={column sep=.65em}},
	grid = both,
	ytick = {0,0.01,0.02,0.03,0.04,0.05,0.06,0.07,0.08},
	]
	
	\addplot[mark=none,color=plotblue] table[x=s,y=ad] {superamplification.dat};
	\addplot[mark=none,color=plotmagenta] table[x=s,y=er] {superamplification.dat};
	\addplot[mark=none,color=plotgreen] table[x=s,y=dep] {superamplification.dat};
	\addplot[mark=none,color=plotorange,dashed] table[x=s,y=ub] {superamplification.dat};
	\legend{$\QC^{(1)}(\NC_s\otimes \AC_{1/2})-\QC^{(1)}(\NC_s)$,$\QC^{(1)}(\NC_s\otimes \EC_{1/2})-\QC^{(1)}(\NC_s)$,$\QC^{(1)}(\NC_s\otimes \DC_{p^*})-\QC^{(1)}(\NC_s)$,Normalized UB on $\QC(\NC_s)$};
	\end{axis}
	\end{tikzpicture}
	\caption{Amplification of coherent information for the channel $\NC_s$ and
        various additional channels.  We plot $\QC^{(1)}(\NC_s\otimes
        \aux)-\QC^{(1)}(\NC_s)$ for $\aux = \EC_{1/2}$ (solid magenta), $\aux =
        \AC_{1/2}$ (solid blue), and $\aux=\DC_{p^*}$ (solid green).  Here,
        $\EC_{1/2}$ and $\AC_{1/2}$ are the symmetric erasure and amplitude
        damping channels respectively, $\DC_{p^*}$ is the qubit depolarizing
        channel with $p^* \approx 0.1893$, so that all  three channels have
        zero coherent information $\QC^{(1)}(\aux)=0$.  We also plot
        $\hat{R}_\alpha(\NC_s)-\QC^{(1)}(\NC_s)$ (dashed orange), where
        $\hat{R}_\alpha(\cdot)$ with $\alpha = 1+2^{-5}$ is the upper bound
        (UB) on the quantum capacity $\QC(\cdot)$ derived in
        \cite{FangFawzi19}.}
	\label{fig:superamp-comparison-main}
\end{figure}
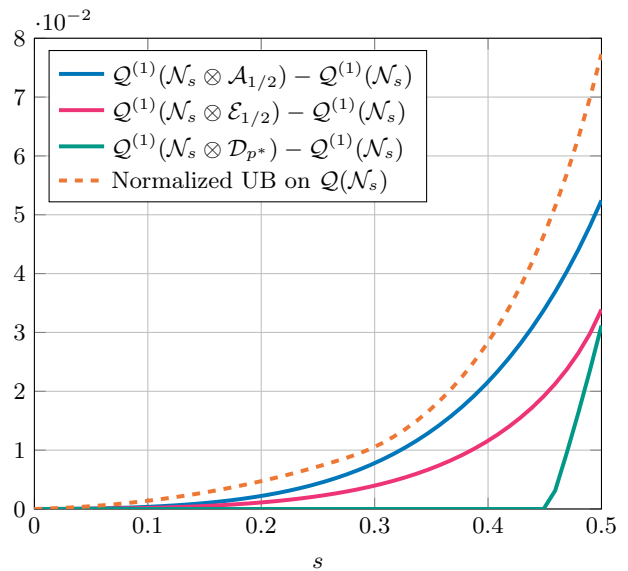

\paragraph*{Violation of strong additivity.}
We find that $\NC_s$ displays super-additivity in the strong sense,
\begin{align}
    \QC^{(1)}(\NC_s \ot \KC) > \QC^{(1)}(\NC_s) + \QC^{(1)}(\KC),
    \label{eq:superAdd}
\end{align}
when used with just about any small channel $\KC$.  Since $\QC^{(1)}(\NC_s)>0$,
the additional channel $\KC$ is said to amplify $\QC^{(1)}(\NC_s)$.
We consider various well-known and physically
relevant channels $\KC$, such as 
the qubit erasure channel, $\EC_{\lm}(\rho) = (1-\lm) \rho + \lm \Tr(\rho)
\dya{e}$ with erasure probability $\lm\in[0,1]$,
the qubit amplitude damping channel, $\AC_{\gm}(\rho) = N_0 \rho N_0^{\dag} +
N_1 \rho N_1^{\dag}$ with damping probability $\gm\in[0,1]$ and Kraus effects
$N_0 = \dya{0} + \sqrt{1-\gm}\dya{1}$ and $N_1 = \sqrt{\gm}\dyad{0}{1}$,
and
the qubit depolarizing channel, $\DC_{p}(\rho) = (1-4p/3) \rho + 2p/3 I$ with
depolarizing parameter $p\in[0,1]$. 
For erasure and amplitude damping channels the quantum capacity equals the channel coherent
information~\cite{BennettDiVincenzoEA97, DevetakShor05, GiovanettiFazio05}.
The amplification in \eqref{eq:superAdd} not only occurs when each of the channels $\EC_{\lm},
\AC_{\gm}$, and $\DC_p$ has zero coherent information~(see
Fig.~\ref{fig:superamp-comparison-main}), but it persists for a wide range of channel
parameters $0 \leq s \leq 1/2$, $\lm_{\min} \leq \lm \leq \lm_{\max}$,
$\gm_{\min} \leq \gm \leq \gm_{\max}$, and $p_{\min} \leq p \leq
p_{\max}$~(see Supplementary material).

Remarkably, the amplification of $\QC^{(1)}(\NC_s)$ by all three channels
$\EC_{\lm}$, $\AC_{\gm}$, and $\DC_p$ can be achieved by a single
input state ansatz for $\NC_s \otimes \KC$, 
\begin{multline}
    \rho(\ep, r_1, r_2) = r_1 \dya{00} + r_2 \dya{01}\\ + (1 - r_1 - r_2) \dya{\chi_{\ep}},
    \label{eq:commInp}
\end{multline}
where $\ket{\chi_{\ep}} = \sqrt{1-\ep} \ket{20} + \sqrt{\ep} \ket{11}$,
and the parameters satisfy the constraints $\ep, r_1, r_2, r_1{+}r_2 \in [0,1]$.
In more detail, we find
that $\Dl^*(\NC_s \ot \KC_x) \coloneqq \max_{\ep, r_1, r_2} \Dl\big(\NC_s \ot \KC_x,
\rho(\ep, r_1, r_2) \big)$ exceeds $\QC^{(1)}(\NC_s)$ + $\QC^{(1)}(\KC_x)$
where $\KC_x$ is one of $\EC_{\lm}$, $\AC_{\gm}$, or $\DC_p$. 
Since all three channels $\EC_{\lm}$, $\AC_{\gm}$, and $\DC_p$ have
well known symmetries, one may suspect that the amplification
strategy~\eqref{eq:commInp} coincides because of these symmetries. 
We find this not to be the case. Numerics reveal that amplification of $\QC^{(1)}(\NC_{1/2})$ using \eqref{eq:commInp} occurs even when $\KC$ is defined in terms of a random qubit channel.
Super-additivity occurs both when $\QC^{(1)}(\KC)>0$ or when the coherent information of $\KC$ itself vanishes.

\paragraph*{Unconditional super-additivity of quantum capacity.}
In the previous section we showed super-additivity of the coherent information of $\NC_s$ when used in parallel with other channels such as $\EC_{\lm}$ or $\AC_{\gm}$.
The latter channels are known to satisfy $\QC(\EC_{\lm}) = \QC^{(1)}(\EC_{\lm})$ and $\QC(\AC_{\gm}) = \QC^{(1)}(\AC_{\gm})$.
Moreover, conditioned on the spin alignment conjecture~(SAC)~\cite{paper1IEEE}, we also have $\QC^{(1)}(\NC_s) = \QC(\NC_s)$.
Hence, the super-additivity of $\QC^{(1)}$ in~\eqref{eq:superAdd} can be elevated to super-additivity of the quantum capacity $\QC$, provided the SAC is true.

We now show that, remarkably, this result can be strengthened to an \emph{unconditional} super-additivity of quantum capacity.
To this end, we consider a channel $\MC_d$ introduced in~\cite{paper1IEEE} that generalizes $\NC_{1/2}$ to $d$ input and output dimensions,
and $d{-}1$ environment dimensions, with $d\geq 3$.  
The isometry $G\colon \HC_a\to \HC_b\otimes \HC_c$ acts on an
orthonormal input basis $\lbrace \ket{i}\rbrace_{i=0}^{d-1}$ as
\begin{align}
\begin{aligned}
		G \ket{0} &= \frac{1}{\sqrt{d-1}}\sum_{j=0}^{d-2} \ket{j} \ot \ket{j}, \\
	G \ket{i} &= \ket{d-1} \ot \ket{i-1} \quad\text{for $i=1,\dots,d-1$},
\end{aligned}
	\label{isoDef2-main}
\end{align}
and defines the channel $\MC_d(\cdot) \coloneqq \tr_c(G\cdot G^\dagger)$.

Comparing \eqref{isoDef2-main} to the isometry \eqref{eq:Ns-iso} for $\NC_{1/2}$, we
see that $\MC_3 = \NC_{1/2}$, and hence $\MC_d$ is indeed a $d$-dimensional
generalization of $\NC_{1/2}$.  The coherent information $\QC^{(1)}(\MC_d)$ is
evaluated in \cite{paper1IEEE}, and similar to $\NC_{1/2}$ we have $\QC(\MC_d) =
\QC^{(1)}(\MC_d)$ modulo (a generalized version of) the spin alignment conjecture.  However, we
do not make use of this (conjectured) identity here and instead use the
following upper bound on the quantum capacity of $\MC_d$ derived in
\cite{paper1IEEE}:
\begin{align}
	\QC(\MC_d) \leq \log\left(1+\frac{1}{\sqrt{d-1}}\right)
	\leq {1 \over \ln 2} \frac{1}{\sqrt{d-1}} \,.\label{eq:Md-Q-bound-main}
\end{align}
This upper bound follows from evaluating the ``transposition bound'' on the quantum capacity of a quantum channel \cite{HolevoWerner01}.
	It is phrased in terms of the diamond norm and can be evaluated using semidefinite programming techniques.

The quantum
capacity of $\MC_{d+1}$ is super-additive when used together with the $d$-dimensional erasure channel
$\EC_{\lm,d}$ where $\lm\in[0,1]$.  
More precisely, we show that
\begin{align}
	\QC(\MC_{d+1} \otimes \EC_{\lm,d}) > \QC(\MC_{d+1}) + \QC(\EC_{\lm,d})\label{eq:Qcap-super}
\end{align}
for suitable $\lambda$ and $d$ in two steps:
First, using the upper bound \eqref{eq:Md-Q-bound-main} on $\QC(\MC_d)$ and
the fact that the quantum capacity of $\EC_{\lm,d}$ is given by
$\QC(\EC_{\lm,d}) = \max\lbrace (1-2\lm)\log d, 0\rbrace$
\cite{BennettDiVincenzoEA97}, we obtain an upper bound
\begin{align}
	u(\lambda,d) \coloneqq \log\left(1+1/\sqrt{d}\right) + \max\lbrace (1-2\lambda)\log d,0\rbrace
    \label{eq:u-ub}
\end{align}
on the right-hand side of \eqref{eq:Qcap-super}.  
Second, letting $\HC_a$ and $\HC_{a'}$ be the input Hilbert spaces for $\MC_{d+1}$ and $\EC_{\lm,d}$, respectively, we find an input state $\rho_{aa'}$ with coherent information $\Delta(\MC_{d+1}\otimes \EC_{\lm,d},\rho_{aa'})$ exceeding $u(\lm,d)$,
\begin{align}
	\QC(\MC_{d+1}) + \QC(\EC_{\lm,d}) &\leq u(\lambda,d)\notag\\
	&< \Delta(\MC_{d+1}\otimes \EC_{\lm,d},\rho_{aa'})\notag\\
	&\leq \QC(\MC_{d+1} \otimes \EC_{\lm,d}).
    \label{eq:Qcap-super-detailed}
\end{align}
This chain of inequalities proves~\eqref{eq:Qcap-super}.

The input state achieving \eqref{eq:Qcap-super-detailed} is $\rho_{aa'} =
\Tr_{rr'}[\psi]_{ara'r'}$, where for $w \in [0,1]$ we define
\begin{multline}
    \ket{\psi}_{ara'r'} = 
    \sqrt{1-w} \; \ket{0}_r\ket{0}_a 
    \; \frac{1}{\sqrt{d}} \sum_{i=0}^{d-1} \ket{i}_{r'} \ket{i}_{a'}\\
	 + \sqrt{w} \; \ket{1}_{r}\ket{0}_{r'} \; \frac{1}{\sqrt{d}} \sum_{i=1}^d \ket{i}_{a} \ket{i-1}_{a'},
	\label{eq:ansatz}
\end{multline}
and the reference spaces $\HC_r$ and $\HC_{r'}$ have dimensions two and $d$, respectively.
The pure state $\ket{\psi}_{ara'r'}$ is a superposition of two orthogonal `pieces` with amplitudes $\sqrt{1-w}$ and $\sqrt{w}$, respectively. 
By itself, the first piece only generates coherent information via $\EC_{\lm,d}$, as the input of $\MC_{d+1}$ in $\HC_a$ is in a product state with both the input to $\EC_{\lm,d}$ and the reference.
The second piece by itself generates no coherent information,
since the joint input system $\HC_{a} \otimes \HC_{a'}$ is unentangled with the reference $\HC_r \otimes \HC_{r'}$.  

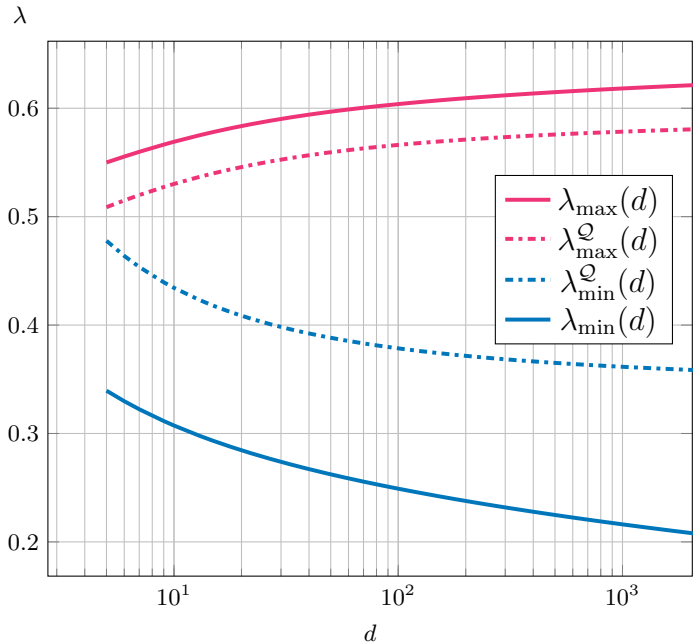
\begin{figure*}[t]
	\centering
	\begin{tikzpicture}
		\begin{axis}[
			xlabel=$d$,
			ylabel=$\lambda$,
			ylabel style={rotate=-90,at = {(0.1,1.05)}},
			scale=1.25,
			every axis plot/.append style={line width=1.5pt},
			xmode = log,
			xmax = 2048,
			legend style={at = {(0.97,0.75)},anchor = north east,font=\large},
			grid = both,
			]
			\addplot[name path = upper,mark=none,color=plotmagenta] table[col sep = comma,x=dExp,y=lmMaxQ1] {MdPlusOneErLmNonAddQ1Log.csv}; %
			\addplot[name path = upper,mark=none,color=plotmagenta,dashdotted] table[col sep = comma,x=dExp,y=lmMaxQExp] {MdPlusOneErLmNonAddQLog.csv};
			\addplot[name path = lower,mark=none,color=plotblue,dashdotted] table[col sep = comma,x=dExp,y=lmMinQExp] {MdPlusOneErLmNonAddQLog.csv};
			\addplot[name path = lower,mark=none,color=plotblue] table[col sep = comma,x=dExp,y=lmMinQ1] {MdPlusOneErLmNonAddQ1Log.csv};%
			\legend{$\lambda_{\mathrm{max}}(d)$,$\lambda_{\mathrm{max}}^{\QC}(d)$,$\lambda_{\mathrm{min}}^{\QC}(d)$,$\lambda_{\mathrm{min}}(d)$};
		\end{axis}
	\end{tikzpicture}
	\caption{
		Plot of the region of super-additivity of coherent information and quantum capacity of the quantum channel $\MC_{d+1} \ot \EC_{\lambda,d}$.
		The solid lines are the minimal values $\lambda_{\mathrm{min}}(d)$ (blue) and maximal values $\lambda_{\mathrm{max}}(d)$ (magenta) between which $\MC_{d+1} \ot \EC_{\lambda,d}$ has super-additive coherent information, $\QC^{(1)}(\MC_{d+1}\otimes \EC_{\lm,d}) > \QC^{(1)}(\MC_{d+1}) + \QC^{(1)}(\EC_{\lm,d})$.
		The dashed lines are the minimal values $\lambda^{\QC}_{\mathrm{min}}(d)$ (blue) and maximal values $\lambda^{\QC}_{\mathrm{max}}(d)$ (magenta) between which $\MC_{d+1} \ot \EC_{\lambda,d}$ has super-additive quantum capacity, $\QC(\MC_{d+1}\otimes \EC_{\lm,d}) > \QC(\MC_{d+1}) + \QC(\EC_{\lm,d})$.
	}
	\label{fig:Q1Q-main}
\end{figure*}

Optimizing over the parameter $w\in[0,1]$, this superposition of coding
strategies results in a coherent information of the joint channel
$\MC_{d+1}\otimes \EC_{\lm,d}$ that exceeds the upper bound $u(\lm,d)$ on
$\QC(\MC_{d+1}) + \QC(\EC_{\lm,d})$.  We first show this numerically for
$\lm\in[0.37,0.57]$ and sufficiently large $d$.  This is summarized in
Fig.~\ref{fig:Q1Q-main}, where we plot the minimal values
$\lambda_{\mathrm{min}}^{\QC}(d)$ (dashed blue) and
$\lambda_{\mathrm{max}}^{\QC}(d)$ (dashed magenta) of $\lambda$ as a function
of $d$ such that \eqref{eq:Qcap-super} holds numerically for all
$\lambda\in[\lambda_{\mathrm{min}}^{\QC}(d),\lambda_{\mathrm{max}}^{\QC}(d)]$.
Note that $\EC_{\lambda,d}$ has positive quantum capacity when $\lambda < 1/2$,
and hence for suitable $d$ and $\lambda$ we obtain super-additivity of
quantum capacity \eqref{eq:Qcap-super} for two channels, $\MC_d$ and $\EC_{\lm,d}$,
each with strictly positive $\QC$.

In Fig.~\ref{fig:Q1Q-main} we also plot the minimal values
$\lambda_{\mathrm{min}}(d)$ (solid blue) and $\lambda_{\mathrm{max}}(d)$ (solid
magenta) such that the coherent information of $\MC_{d+1}\otimes \EC_{\lm,d}$
is super-additive for all
$\lambda\in[\lambda_{\mathrm{min}}(d),\lambda_{\mathrm{max}}(d)]$. %
While the interval
$[\lambda_{\mathrm{min}}(d),\lambda_{\mathrm{max}}(d)]$ marks the `true' extent
of the super-additivity of quantum capacity (modulo the spin alignment
conjecture), we stress once again that the super-additivity of quantum capacity
within the interval
$[\lambda_{\mathrm{min}}^{\QC}(d),\lambda_{\mathrm{max}}^{\QC}(d)]$ is
unconditional.

We can further strengthen the numerical results of Fig.~\ref{fig:Q1Q-main} by
proving analytically that the super-additivity of quantum capacity in
\eqref{eq:Qcap-super} indeed holds for all $\lambda\in(0,1)$ and sufficiently
large $d$.  The proof is based on a log-singularity-like argument \cite{Siddhu21}, 
and applied for any $\lambda\in(0,1)$, by a suitable choice of the
parameter $w$ in the state \eqref{eq:ansatz}.  Details of this calculation
can be found in the Supplementary material.

\paragraph*{Discussion.}
Interestingly, a single ansatz~\eqref{eq:ansatz} is responsible for
super-additivity of $\QC^{(1)}$ when $\NC_s$ is used with a variety of other
channels $\EC_{\lm}, \AC_{\gm}, \DC_{p}$, and randomly constructed qubit
channels.  A higher dimensional version of this ansatz gives rise to
super-additivity of quantum capacity when $\MC_d$ is used with $\EC_{\lm,d}$.
The mechanism and extent of this super-additivity is distinct from
super-activation, where the private capacity of a zero quantum capacity channel
$\NC$ is transformed into quantum capacity when used jointly with an
anti-degradable channel $\AC$.  This transformation has efficiency at most
$1/2$, and thus one obtains super-activation when $ 0 = \QC(\NC) < \PC(\NC)/2$.
By contrast, $\QC(\NC_s) > \PC(\NC_s)/2 > 0$, thus ruling out the
super-activation mechanism as the cause for our super-additivity involving
$\NC_s$; our protocol~\eqref{eq:ansatz} employs a different mechanism.

Like super-activation our protocol works robustly~\cite{BrandaoOppenheimEA12}
when $\AC = \EC_{\lm,d}$ and $\lm$ is varied, but unlike super-activation we
find super-amplification, i.e., super-additivity even when both channels
$\MC_d$ and $\EC_{\lm,d}$ have non-zero quantum capacity.  Similar
super-additivity of quantum capacity arises in high-dimensional rocket and
half-rocket channels when used with zero capacity
channels~\cite{SmithSmolin09a,LeungLiEA14}. These noisy channels, carefully
constructed to display super-additivity, have quantum capacity well below the
dimensional bound $r = \QC/ \log d \ll 1$.  By contrast, $\MC_d$ is simply
constructed by hybridizing a degradable qubit channel with a useless channel,
with the goal to support weak additivity of $\QC^{(1)}$.  Yet, it exhibits
super-additivity of $\QC$ even when it has modest input dimension and noise;
for instance super-additivity occurs at $d = 5$ and $r > .2$.  Our result on
$\MC_d$ also contrasts with those obtainable by extending super-activation via
continuity arguments.  The super-activating channels can be perturbed to have
positive capacities, but these capacities are necessarily very small.
Moreover, super-additivity involving $\MC_d$ occurs over a wide range of
erasure probabilities that is well beyond what one may expect from such
perturbations.  For instance, at $d = 10$, $r \simeq .075$, and
super-additivity holds over erasure probabilities $.43 \leq \lambda \leq .53$,
and the erasure channel can have substantial capacity. 
Using $\MC_d$ with a symmetric channel, $\SC$, of unbounded dimension leads to
super-additivity, $\QC(\MC_d \ot \SC) \geq \QC(\MC_d) + \QC(\SC)$ for any $d
\geq 7$ where $\PC(\MC_d)/2> \QC(\MC_d)$~\cite{paper1IEEE}, since $\QC(\MC_d \ot \SC) >
\PC(\MC_d)/2$~\cite{SmithYard08}.
These super-additivity results can be strengthened and simplified further if
the SAC is proven.  The simplicity of the channels involved in super-additivity
here raises the question of whether  qualitatively similar constructions are
possible for investigating super-additivity of private and classical
capacities.

\paragraph*{Acknowledgments.}
This work was partially supported by ARO MURI Quantum Network Science under
contract number W911NF2120214, NSF grants CCF 1652560, PHY 1915407, 2137953 and
an NSERC discovery grant.

\clearpage

%

\clearpage
\onecolumngrid

\section*{\texorpdfstring{Supplementary information for\\[0.5em]``Generic nonadditivity of quantum capacity in simple channels''}{Supplementary information}}\label{supp}

Here we provide additional information about the results discussed in the main text.
We start in Sec.~\ref{sec:prelim} with preliminaries on quantum channels and their quantum capacity.
We then discuss violations of strong additivity of quantum capacity involving the channels $\NC_s$ and $\MC_d$ in Secs.~\ref{sec:supAmp} and~\ref{sec:amplification-Md}, respectively.

\section{Preliminaries}
\label{sec:prelim}
\subsection{Quantum channels}\label{sec:capacities}

Let $\HC$ be a Hilbert space of finite dimension $d$.
Let $\HC^{\dag}$ be the dual of $\HC$, and
$\hat \HC \cong \HC \ot \HC^{\dag}$ be the space of linear operators
acting on $\HC$.
Throughout this paper, the ket $|\psi\rangle \in \HC$ denotes a unit
vector in $\HC$, and the bra $\langle \psi| \in \HC^{\dag}$ is the
dual vector.
We use $[\psi]$ as a shorthand for $|\psi\rangle \langle \psi|$.

Let $\HC_a, \HC_b,$ and $\HC_c$ be three Hilbert
spaces of dimensions $d_a, d_b,$ and $d_c$ respectively.  An isometry $E:\HC_a
\mapsto \HC_b \ot \HC_c$, i.e., a map satisfying $E^{\dag} E = I_a$~(the
identity on $\HC_a$), takes an input Hilbert space $\HC_a$ to a subspace of a
pair of output spaces $\HC_b \ot \HC_c$.  This isometry generates a quantum
channel pair, $(\BC, \BC^c)$, i.e., a pair of completely positive~(CP) trace-preserving~(TP) maps, with superoperators,
\begin{align}
	\BC(X) = \Tr_c(EXE^{\dag}), \quad \text{and} \quad
	\BC^c(X) = \Tr_b(EXE^{\dag}),
	\label{eq:chanPair}
\end{align}
that take any element $X \in \hat \HC_a$ to $\hat \HC_b$ and $\hat \HC_c$,
respectively.  Each channel in this pair $(\BC,\BC^c)$ may be called the
complement of the other. If the input of the isometry $E$ is restricted to a
subspace $\HC_{\bar a}$ of $\HC_a$, then such a restricted map is still an
isometry on $\HC_{\bar a}$ and defines a pair of channels $(\bar \BC, \bar
\CC)$, where each channel $\bar \BC$ and $\bar \CC$ is called a {\em
	sub-channel} of $\BC$ and $\CC$, respectively.  When focussing on some quantum
channel $\BC$, it is common to refer to $\HC_a,\HC_b,$ and $\HC_c$ as the channel
input, output, and environment respectively.  

Any CPTP map~(together with its complement) may be written as~\eqref{eq:chanPair} in terms of a suitable
isometry $E$.  
Another representation of a CPTP map comes from its Choi-Jamio\l{}kowski
operator.  To define this operator, consider a linear map $\BC \colon \hat
\HC_a \mapsto \hat \HC_b$, an orthonormal basis $\{\ket{i}_{a}\}$ on
$\HC_a$, and a maximally entangled state,
\begin{align}
	\ket{\phi} = \frac{1}{\sqrt{d_a}} \sum_{i=1}^{d_a} \ket{i}_{a} \ot \ket{i}_{a},
	\label{eq:maxEnt}
\end{align}
on $\HC_{a} \ot \HC_a$. The {\em unnormalized} Choi-Jamio\l{}kowski operator
of $\BC$ is
\begin{align}
	J^{\BC}_{ab} = d_a(\IC_{a} \ot \BC)  (\dya{\phi})\,,
	\label{eq:cjOp}
\end{align}
where $\IC_a$ denotes the identity map acting on $\hat \HC_a$.  The linear map
$\BC$ is CP if and only if the above operator is positive semidefinite, and TP if and only if
its partial trace over $\HC_b$ is the identity $I_{a}$ on $\HC_{a'}$.

\subsection{Quantum capacity}

The quantum capacity $\QC(\BC)$ of a quantum channel
$\BC\colon\hat{\HC_a}\to\hat{\HC_b}$ is defined as the largest rate at which
quantum information can be sent faithfully through the channel.  It can be
expressed in terms of an entropic quantity as follows.  Let $\rho_a$ denote a
density operator~(unit trace positive semi-definite operator) on $\HC_a$ and
for any $\rho_a$ let $\rho_b \coloneqq \BC(\rho_a)$ and $\rho_c \coloneqq
\BC^c(\rho_a)$.  The coherent information~(or entropy bias) of a channel $\BC$
at a density operator $\rho_a$ is
\begin{align}
	\Dl(\BC, \rho_a) = S(\rho_b) - S(\rho_c),
	\label{entBias}
\end{align}
where $S(\rho) = -\Tr(\rho \log \rho)$~(we use $\log$ base $2$ by default) is
the von-Neumann entropy of $\rho$. The channel coherent information~(sometimes
called the {\em single-letter} coherent information),
\begin{align}
	\QC^{(1)}(\BC) = \max_{\rho_a} \Dl(\BC, \rho_a),
	\label{eq:q1Deg}
\end{align}
is an achievable rate for sending quantum information across the
channel $\BC$, and hence $\QC(\BC)\geq
\QC^{(1)}(\BC)$~\cite{Lloyd97,Shor02a,Devetak05}. The maximum achievable rate
is equal to the quantum capacity of $\BC$, and given by a {\em multi-letter}
formula~(sometimes called a regularized
expression)~\cite{BarnumNielsenEA98,BarnumKnillEA00,Lloyd97,Shor02a,Devetak05},
\begin{align}
	\QC(\BC) = \sup_{n \in \mathbb{N}} \frac{1}{n} \QC^{(1)}(\BC^{\ot n}),
	\label{eq:chanCap}
\end{align}
where $\BC^{\ot n}$ represent $n \in \mathbb{N}$ parallel~(sometimes called
joint) uses of $\BC$.  
In contrast, \eqref{eq:q1Deg} is often called the ``single-letter expression.''

\subsection{Superadditivity, amplification and super-amplification}
\label{sec:ampframework}

The channel coherent information is {\em super-additive}.  For any two
quantum channels $\BC$ and $\BC'$ used together, the channel coherent
information of the joint channel $\BC \ot \BC'$ satisfies an
inequality,
\begin{align}
	\QC^{(1)}(\BC \ot \BC') \geq \QC^{(1)}(\BC) + \QC^{(1)}(\BC'),
	\label{eq:nonAddCI}
\end{align}
which is known to be strict for some $\BC$ and $\BC'$~\cite{DiVincenzoShorEA98, FernWhaley08, SmithSmolin07, SmithYard08, SmithSmolin09a, LeditzkyLeungEA18, BauschLeditzky19, BauschLeditzky20, SiddhuGriffiths20, Siddhu20, Siddhu21}. 
We will use the following terminology for special cases of super-additivity to facilitate our discussion.

We say that the coherent information of a channel $\BC$ is {\em weakly super-additive} if there exists $n \in \mathbb{N}$ such that 
\begin{align}
	\frac{1}{n} \QC^{(1)}(\BC^{\ot n})
	> \QC^{(1)}(\BC) \,.\label{eq:CI-super-add}
\end{align}
In this case the channel capacity $\QC(\BC)$, equal to the supremum of the left-hand side of \eqref{eq:CI-super-add} over all $n \in \mathbb{N}$, is strictly greater than
the channel coherent information, and hence the regularization
in~\eqref{eq:chanCap} is necessary~\cite{DiVincenzoShorEA98, FernWhaley08, SmithSmolin07, LeditzkyLeungEA18a, BauschLeditzky19, BauschLeditzky20, SiddhuGriffiths20, Siddhu20, Siddhu21}.

If a strict inequality holds in \eqref{eq:nonAddCI} for two \emph{different} channels $\BC$ and $\BC'$, we say that the coherent
information is {\em strongly super-additive}. 
It is known that the quantum capacity can be strongly super-additive as well, i.e., there are channels $\BC$ and $\BC'$ such that \cite{SmithYard08}
\begin{align}
	\QC(\BC \otimes \BC') > \QC(\BC) + \QC(\BC').\label{eq:Qcap-super-add} 
\end{align}

Due to the potential need for regularization for evaluating the quantum capacity, super-additivity has typically been demonstrated for channels with vanishing or well-bounded capacities in order to certify a strict inequality in 
\eqref{eq:Qcap-super-add}.
A particular example of this is the case $\QC(\BC)=\QC(\BC')=0$, which is
called {\em superactivation}~\cite{SmithYard08}.
If $\QC(\BC)>0$ and $\QC(\BC')=0$, %
we call the super-additivity of quantum capacity {\em amplification}.  
If $\QC(\BC)>0$ and $\QC(\BC')>0$,  
we call the super-additivity of quantum capacity {\em super-amplification}.  
We use similar terminology for the channel coherent information $\QC^{(1)}(\cdot)$ (defined in \eqref{eq:q1Deg}) of a quantum channel.
The main results of this paper are demonstrations of both amplification and super-amplification of coherent information and quantum capacity.

\subsection{Special channel classes}
\label{sec:spChanCl}

A channel $\BC$ is called {\em degradable}, and its complement $\BC^c$ {\em
	anti-degradable}, if there is another channel $\DC$ such that $\DC \circ \BC =
\BC^c$~\cite{DevetakShor05,CubittRuskaiEA08}. Sometimes this channel $\DC$ is
called the {\em degrading map} of the degradable channel $\BC$. For any two
channels $\BC'$ and $\BC$, each either degradable or anti-degradable, the joint
channel $\BC \ot \BC'$ has additive coherent information.
For a degradable channel $\BC$, the coherent information $\Dl(\BC, \rho_a)$ is
concave in $\rho_a$~\cite{YardHaydenEA08}, and thus $\QC^{(1)}(\BC)$ can be
computed with relative ease~\cite{FawziFawzi18, RamakrishnanItenEA21}. As a
result the quantum capacity of a degradable channel, which simply equals
$\QC^{(1)}(\BC)$, can also be computed efficiently.  An anti-degradable channel has no
quantum capacity due to the no-cloning theorem.  

Besides anti-degradable channels, the only other known class of zero-quantum-capacity
channels are {\em entanglement binding} or {\em positive under
	partial-transpose}~(PPT) channels~\cite{HorodeckiHorodeckiEA00}.  A channel is
PPT if its Choi-Jamio\l{}kowski operator~\eqref{eq:cjOp} is positive under
partial transpose.

\section{Violation of strong additivity of quantum capacity involving $\NC_s$}
\label{sec:supAmp}

\subsection{The $\NC_s$ channel}
\label{sec:NsChan}

Let $\HC_a, \HC_b,$ and $\HC_c$ have dimensions $d_a = d_b = 3$, and $d_c=2$.
Consider an isometry $F_s\colon \HC_a \mapsto \HC_b \ot \HC_c$ with $0 \leq s \leq 1/2$ of the form
\begin{align}
	F_s \ket{0} &= \sqrt{s} \; \ket{0} \ot \ket{0} + \sqrt{1-s} \; \ket{1} \ot \ket{1},  \nonumber \\
	F_s \ket{1} &= \ket{2} \ot \ket{0}, \nonumber \\
	F_s \ket{2} &= \ket{2} \ot \ket{1}.
	\label{isoDef1}
\end{align}
This isometry was introduced previously by one of us in~\cite{Siddhu21} with
$\ket{1}$ and $\ket{2}$ in $\HC_a$ exchanged. 
Furthermore, the channel defined via \eqref{isoDef1} is unitarily equivalent to a quantum channel introduced in \cite{WangDuan18} and further studied in \cite{WangXieEA17} (see \cite{paper1IEEE} for a more detailed discussion).
The isometry~\eqref{isoDef1}
gives rise to a complementary pair of channels $\NC_s\colon \hat \HC_a \mapsto
\hat \HC_b$ and $\NC^c_s \colon \hat \HC_a \mapsto \hat \HC_c$. 
The channel maps an input operator $\rho = \sum_{ij} \rho_{ij} \dyad{i}{j}$
to
\begin{equation}
    \NC_s(\rho) = 
\begin{pmatrix}
    s \rho_{00} & 0 & \sqrt{s} \rho_{01} \\
    0 & (1-s)\rho_{00} & \sqrt{1-s} \rho_{02} \\
    \sqrt{s} \rho_{10} & \sqrt{1-s} \rho_{20} & \rho_{11} + \rho_{22} \\
\end{pmatrix}.
    \label{eq:rhoOut}
\end{equation}

In \cite{paper1IEEE}, the capacities of $\NC_s$ and $\NC_s^c$ are studied
in detail.  It is proved that both $\NC_s$ and $\NC_s^c$ are neither
degradable nor antidegradable, and neither channel belongs to any class of
channels with known quantum, private, or classical capacity.
Surprisingly, the capacities can still be found through a variety
of techniques (see \cite{paper1IEEE} for details). In summary, the
quantum, private, and classical capacities of $\NC_s^c$ are all
equal to $1$: 
\begin{align}
	\QC(\NC_s^c)  = \PC(\NC_s^c) = \CC(\NC_s^c) = 1.
	\label{eq:NsCompCap}
\end{align}
Proof for these equalities relies on identifying a perfect channel from
the two-dimensional input subspace $\HC_a$, spanned by $\{ \ket{1}, \ket{2}\}$,
to the two dimensional output $\HC_c$ of $\NC_s^c$.
The private and classical capacity of $\NC_s$ likewise are equal to $1$, and
the underlying information quantities turn out to be additive:
$\PC^{(1)}(\NC_s) = \chi(\NC_s) = \PC(\NC_s) = \CC(\NC_s) = 1$. 
Proof for these equalities shows a lower bound on $\PC^{(1)}(\NC_s)$
that matches an upper bound on $\CC(\NC_s)$. The lower bound on
$\PC^{(1)}(\NC_s)$ uses an explicit ensemble of two orthogonal inputs that
remain orthogonal at the channel output $\HC_b$ but become indistinguishable at
the complementary output $\HC_c$. The upper bound on $\CC(\NC_s)$ uses an
explicit solution to a semidefinite programming upper bound on $\CC$ proved in~\cite{WangXieEA17}.
The coherent information of $\NC_s$ is the solution of a one-parameter concave
maximization problem over a bounded interval. For any $0 \leq s \leq 1/2$,
\begin{align}
	\QC^{(1)}(\NC_s) = \max_{\rho_a(u)} \Dl\big( \NC_s, \rho_a(u) \big),
	\label{eq:NsQ1}
\end{align}
where $\rho_a(u)$ is a one-parameter density operator of the form $\rho_a(u) =
(1-u)[0] + u [2]$ with $0 \leq u \leq 1$.
Moreover, provided the spin alignment conjecture stated in \cite{paper1IEEE} is true,
the coherent information $\QC^{(1)}(\NC_s)$ is also additive, and
$\QC^{(1)}(\NC_s) = \QC(\NC_s)$.
Proof for this equality has two steps. First step identifies that an
input restricted to a degradable sub-channel of $\NC_s$ achieves the
single-letter coherent information $\QC^{(1)}(\NC_s)$. 
The second step shows that inputs restricted to tensor products of this
degradable sub-channel achieve the multi-letter coherent information of $\NC_s$
conditioned on what we call the spin-alignment conjecture.
Together, the results in \cite{paper1IEEE} show that the coherent information, the
private information, and the Holevo information of $\NC_s$ are all
weakly additive.

Operationally, a channel having a weakly additive information quantity means that a more complex coding strategy (defined by typical subspaces of multi-letter entangled input states) does not increase the rate. 
It is hence surprising that, for coherent information, the qutrit channel $\NC_s$
exhibits strong super-additivity when it is used jointly with a {\em
	different} channel $\aux$.  Even more surprisingly, this additional
channel $\aux$ can be just about any small channel!  
For example, we found super-additivity when $\aux$ is chosen from well-known and physically relevant channels, such as the amplitude damping channel, erasure channel, and depolarizing channel on a qubit, over large ranges of the corresponding noise parameters.
Alternatively, this additional channel can be randomly sampled and a substantial fraction
have super-additive coherent information jointly with $\NC_{1/2}$.
Note that $\QC(\NC_s)>0$ for all $s\in[0,1/2]$ (see Section~4 and Fig.~1 in \cite{paper1IEEE}), and hence the additional channel $\aux$ \emph{amplifies} the coherent information of $\NC_s$ (in the terminology introduced in Sec.~\ref{sec:ampframework} above):
\begin{align}
	\QC^{(1)}(\NC_s \otimes \aux) > \QC^{(1)}(\NC_s) + \QC^{(1)}(\aux).\label{eq:superamplification}
\end{align}
Finally, as many of the additional channels have known quantum
capacity, some positive and some vanishing, the inequality \eqref{eq:superamplification} can be lifted to
super-amplification and amplification of quantum capacity subject to the spin
alignment conjecture \cite{paper1IEEE}.

We will describe the aforementioned results in detail in the rest of
this section.  The effect can be demonstrated for a large range
of $s$ and many choices of $\aux$; to facilitate the discussion, we
organize the presentation as follows.
In Section~\ref{sec:depSymm} we focus on three specific choices of
$\aux$ that have vanishing coherent information
and demonstrate amplification of coherent
information with $\NC_s$ for a large range of the parameter $s$.
In Section~\ref{sec:erasure}, we instead fix $s=1/2$ and use the qubit erasure channel as the additional channel $\aux$, varying the noise parameter of the latter over the full range.
We show amplification of coherent information in a very large interval of this noise parameter, including regimes when $\QC^{(1)}(\aux)=\QC(\aux)>0$, hence demonstrating super-amplification.
We examine the novel mechanism behind the (super-)amplification.
In Section~\ref{sec:superNs}, we return to the full family of $\NC_s$,
and three single-parameter families of $\aux$ that extend the examples
in Section~\ref{sec:depSymm}, and present large regions of joint noise
paramters for (super-)amplification.
Furthermore, two of our extended families of $\aux$ have quantum
capacities equal to the coherent information.  Therefore, conditioned
on the spin alignment conjecture, all aforementioned results
on (super-)amplification hold also for quantum capacity.
In Section~\ref{sec:randchan}, we describe the prevalence of
amplification and super-amplification over randomly chosen $\aux$.

\subsection{Amplification by channels with zero coherent information or quantum capacity}
\label{sec:depSymm}

In this section, we illustrate amplification in \eqref{eq:superamplification} by choosing the additional channel $\aux$ to be
the erasure channel $\EC_\lambda$,
the amplitude damping channel $\AC_\gamma$,
and the depolarizing channel $\DC_p$. 
In each case, a non-negative parameter $x$ determines the noise level; in this section, we fix the noise level $x$ for each channel so that the coherent information vanishes, $\QC^{(1)}(\aux_x)=0$.
Since the erasure channel and the amplitude damping channel are either degradable or antidegradable for all values of the noise level $x$, their coherent information is additive and equal to the quantum capacity, $\QC^{(1)}(\aux_x) = \QC(\aux_x)$.
Hence, in these cases we also have $\QC(\aux_x) = 0$, which illustrates amplification of quantum capacity subject to the spin alignment conjecture.

\begin{figure}[ht!]
	\centering
	\begin{tikzpicture}
		\begin{axis}[
			xlabel=$s$,
			xmin = 0,
			xmax = 0.5,
			ymin = 0,
			ymax = 0.08,
			scale=1.25,
			every axis plot/.append style={line width=1.5pt},
			legend cell align={left},
			legend style={at = {(0.025,0.975)},anchor = north west,/tikz/every even column/.append style={column sep=.65em}},
			grid = both,
			ytick = {0,0.01,0.02,0.03,0.04,0.05,0.06,0.07,0.08},
			]
			
			\addplot[mark=none,color=plotblue] table[x=s,y=ad] {superamplification.dat};
			\addplot[mark=none,color=plotmagenta] table[x=s,y=er] {superamplification.dat};
			\addplot[mark=none,color=plotgreen] table[x=s,y=dep] {superamplification.dat};
			\addplot[mark=none,color=plotorange,dashed] table[x=s,y=ub] {superamplification.dat};
			\legend{$\QC^{(1)}(\NC_s\otimes \AC_{1/2})-\QC^{(1)}(\NC_s)$,$\QC^{(1)}(\NC_s\otimes \EC_{1/2})-\QC^{(1)}(\NC_s)$,$\QC^{(1)}(\NC_s\otimes \DC_{p^*})-\QC^{(1)}(\NC_s)$,Normalized UB on $\QC(\NC_s)$ from \cite{FangFawzi19}};
		\end{axis}
	\end{tikzpicture}
	\caption{
		Amplification of coherent information for the channel $\NC_s$ and various additional channels.
		We plot $\QC^{(1)}(\NC_s\otimes \aux)-\QC^{(1)}(\NC_s)$ for $\aux = \EC_{1/2}$ (solid magenta), 
		$\aux = \AC_{1/2}$ (solid blue), and $\aux=\DC_{p^*}$ (solid green).  Here, $\EC_{1/2}$ and $\AC_{1/2}$
		are the symmetric erasure and amplitude damping channels respectively, 
		$\DC_{p^*}$ is the qubit depolarizing channel with $p^* \approx 0.1893$, so that all  
		three channels have zero coherent information $\QC^{(1)}(\aux)=0$.  
		We also plot $\hat{R}_\alpha(\NC_s)-\QC^{(1)}(\NC_s)$ (dashed orange), where $\hat{R}_\alpha(\cdot)$ with $\alpha = 1+2^{-5}$ is the upper bound (UB) on the quantum capacity $\QC(\cdot)$ derived in \cite{FangFawzi19}.
		This figure is identical to Fig.~1 in the main text and reproduced here for convenience.
	}
	\label{fig:superamp-comparison}
\end{figure}
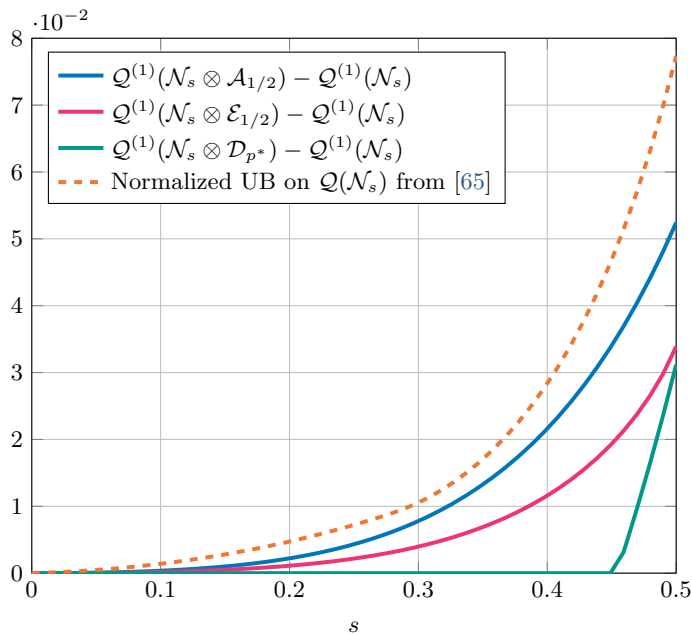

We now describe the assisting channels in more detail, and compute both sides of \eqref{eq:superamplification} numerically for the whole interval $s\in[0,1/2]$ with the three chosen additional channels.
These numerical results are summarized in Fig.~\ref{fig:superamp-comparison}.
We also present a {\em common} structure of the joint input state 
achieving \eqref{eq:superamplification}.

\paragraph{Erasure channel.}
The general erasure channel has input space $\HC_{a'}$ and output and environment spaces $\HC_{b'} = \HC_{c'} = \HC_{a'} \oplus \mathbb{C} |e\rangle$, where $|e\rangle$ is an `erasure flag' orthogonal to the input space $\HC_{a'}$.
The erasure channel acts as follows: with erasure probability
$\lambda$ the channel erases the input and replaces it with the
erasure flag $[e]$, while with probability $1-\lambda$ the
channel transmits the input unaltered.
When $\HC_{a'}$ is $2$-dimensional, the erasure channel pair $(\EC_\lambda,\EC_\lambda^c)$ can be generated from a channel isometry $E_\lambda$ defined via
\begin{align}
	E_\lambda\, |0\rangle &= \sqrt{1-\lambda}\, |0e\rangle + \sqrt{\lambda} \, |e0\rangle \,, & E_\lambda\, |1\rangle &= \sqrt{1-\lambda}\, |1e\rangle + \sqrt{\lambda} \, |e1\rangle \,. \label{eq:ErIso} 
\end{align}
Note that the complementary channel $\EC^c_\lambda$ is an erasure channel with erasure probability $1-\lambda$, i.e., $\EC^c_\lambda = \EC_{1-\lambda}$, and the erasure channel is degradable for $\lambda\in[0,1/2]$ and anti-degradable for $\lambda\in[1/2,1]$.
As a result, the channel coherent information coincides with the quantum capacity for all $\lambda$, taking on the simple form~\cite{BennettDiVincenzoEA97}
\begin{align}
	\QC^{(1)}(\EC_{\lm}) = \QC(\EC_\lambda) = \max\lbrace 1-2\lambda,0\rbrace.
	\label{eq:ErQ1Q}
\end{align}

We fix $\lambda = 1/2$ so that $\QC^{(1)}(\EC_{\lm}) = \QC(\EC_{1/2}) = 0$ and we numerically maximize the coherent information $\QC^{(1)}(\NC_s\otimes \EC_{1/2})$ for $s\in[0,1/2]$.
We find that its value exceeds $\QC^{(1)}(\NC_s)$ for all $s\in(0,1/2)$, as shown by the solid magenta line in Fig.~\ref{fig:superamp-comparison}.
An exhaustive search
over the full input space reveals the following structure of the joint input $\rho_{aa'}$ that attains the maximum joint coherent information:
\begin{align}
	\rho_{aa'} \equiv \rho_{aa'}(\epsilon,r_1,r_2) = r_1 [00]_{aa'} + r_2 [01]_{aa'} + (1-r_1-r_2) [\chi_\epsilon]_{aa'},\label{eq:ampNsAnz}
\end{align}
where $\HC_a$ and $\HC_{a'}$ denote the input spaces of $\NC_s$ and $\EC_{1/2}$, respectively, 
$|\chi_\epsilon\rangle_{aa'} = \sqrt{1-\epsilon} \, |20\rangle_{aa'} + \sqrt{\epsilon} \, |11\rangle$,
and the parameters $r_1,r_2,\epsilon\in[0,1]$, $r_1+r_2\leq 1$ are optimized.
In Section~\ref{sec:erasure} we further analyze the coherent information for $\NC_{1/2}\otimes\EC_\lambda$ for varying $\lambda\in[0,1]$ and discuss the mechanism for the amplification.

\paragraph{Amplitude damping channel.}

The amplitude damping channel is defined as follows.
Let $\HC_{a'}, \HC_{b'},$ and $\HC_{c'}$ be two-dimensional Hilbert spaces.
Consider an isometry $A_\gamma\colon \HC_{a'} \mapsto \HC_{b'} \ot \HC_{c'}$ of the
form
\begin{align}
	A_\gamma \, \ket{0} &= \ket{00} \,, & A_\gamma \, \ket{1}
	&= \sqrt{\gamma} \, \ket{01} + \sqrt{1-\gamma} \, \ket{10}\,.
	\label{eq:isoAmpD}
\end{align}
The isometry above defines a channel pair $(\AC_\gamma, \AC_{1-\gamma})$, where $\AC_\gamma \colon \hat \HC_{a'} \mapsto \hat \HC_{b'}$ is a qubit amplitude damping channel with damping probability $\gamma$, and $\AC_\gamma^c = \AC_{1-\gamma}$. When $\gamma = 1/2$,
$\AC_\gamma = \AC_\gamma^c$ so the two channels are symmetric.  
The channel coherent information is attained on
$\rho(z) = (1-z) [0] + z [1]$ \cite{GiovanettiFazio05}, and hence 
\begin{align}
	\QC^{(1)}(\AC_\gamma) = \max_{0 \leq z \leq 1} \Dl(\AC_\gamma, \rho(z)) \,. 
	\label{eq:q1Acp}
\end{align}
The channel $\AC_\gamma$ is degradable for $\gamma\in[0,1/2]$, and
anti-degradable for $\gamma \in[1/2,1]$.
As a result, for all $\gamma \in[0,1]$
the coherent information \eqref{eq:q1Acp} is additive, 
\begin{align}
	\QC(\AC_\gamma) = \QC^{(1)}(\AC_\gamma) \,.
	\label{eq:AmpDAdd}
\end{align}
This capacity is zero in the antidegradability regime, $\QC(\AC_\gamma) = 0$ for $\gamma \geq 1/2$.  Similar to the earlier analysis involving the erasure channel, we choose $\gamma=1/2$, and numerically maximize $\QC^{(1)}(\NC_s\otimes \AC_{1/2})$.  
Again, we find that $\QC^{(1)}(\NC_s\otimes \AC_{1/2}) > \QC^{(1)}(\NC_s)$ for all $s\in(0,1/2)$, as shown by the solid magenta line in Fig.~\ref{fig:superamp-comparison}.
The optimal input
achieving the maximal coherent information has the same form given
by~\eqref{eq:ampNsAnz}, just as in the case when $\NC_s$ is used with
the erasure channel.  

\paragraph{Depolarizing channel.}
The qubit depolarizing channel has input and output space $\HC_{a'} = \HC_{b'} = \mathbb{C}^2$ and environment space $\HC_{c'} = \mathbb{C}^4$.
The depolarizing channel acts as follows: with probability $4p/3$ the channel replaces the input by the maximally mixed state $I_a/2$, or in Kraus representation, 
\begin{align}
	\DC_p(\rho) = (1-p) \, \rho + \frac{p}{3}\left(X\rho X + Y\rho Y + Z\rho Z\right)\,,
	\label{eq:Dp}
\end{align}
where $0 \leq p \leq 3/4$ and $X,Y,Z$ are the Pauli matrices.
The depolarizing channel is anti-degradable for $p\geq 1/4$, and hence $\QC(\DC_p) = 0$ in that regime.
For $p\in(0,1/4)$, the quantum capacity of $\DC_p$ is unknown, and only lower and upper bounds on $\QC(\DC_p)$ are available \cite{DiVincenzoShorEA98,SmithSmolin07,FernWhaley08,BauschLeditzky20,BauschLeditzky19,LeditzkyDattaEA18,LeditzkyLeungEA18a,wang2021pursuing,FanizzaKianvashEA20,kianvash2022bounding}.
For our discussion, we fix $p^*=0.1893$, which is known as the ``hashing point'', defined to be the smallest $p$ such that $\QC^{(1)}(\DC_p) = 0$.
Note that, in contrast to the additional channels used earlier, $\DC_{p^*}$ does have positive quantum capacity, $\QC(\DC_{p^*})>0$ \cite{ShorJohn96,DiVincenzoShorEA98}, but we do not know its exact value.

The numerically optimized coherent information of the joint channel $\NC_{s}\otimes \DC_{p^*}$ is presented in Fig.~\ref{fig:superamp-comparison}, with the solid green line showing that $\QC^{(1)}(\NC_s\otimes \DC_{p^*}) > \QC^{(1)}(\NC_s)$ for $s\in[0.45,0.5]$.
The joint input achieving the maximal coherent information for $\NC_s\otimes \DC_{p^*}$ again has the same form~\eqref{eq:ampNsAnz} as the optimal states for the qubit amplitude damping and qubit erasure channel. 

\subsection{(Super-)amplification of coherent information of $\NC_{1/2}$
	and erasure channel}
\label{sec:erasure}

In this subsection, we fix $s=1/2$ and consider the joint optimization
in $\QC^{(1)}(\NC_{1/2} \otimes \aux_x)$ for $\aux_x$ being the
continuous families of erasure and amplitude damping channels.  We
find both amplification and super-amplification of coherent
information over large ranges of the noise parameter $x$.  As $\aux_x$
has additive coherent information, our results also imply
amplification and super-amplification of quantum capacity, conditioned
on the spin alignment conjecture.  We focus on 
$\NC_{1/2}$ used jointly with the erasure channel, since this allows for a
better comparison with earlier results on super-additivity of quantum
capacity.  However, we note that similar results also hold for
the amplitude damping channel.

\begin{figure}[h]
	\centering
	\begin{tikzpicture}
		\begin{axis}[
			xlabel=$\lm$,
			xmin = 0.3,
			xmax = 1,
			ymin = -0.01,
			ymax = 0.04,
			scale=1.5,
			every axis plot/.append style={line width=1.5pt},
			legend cell align={left},
			legend style={at = {(0.5,1.1)},anchor = south,/tikz/every even column/.append style={column sep=.65em}},
			grid = both,
			extra x ticks = {0.41,0.593,0.723},
			extra x tick labels = {$\lm_{\mathrm{min}}$,$\lm_r$,$\lm_{\mathrm{max}}$},
			extra x tick style = {ticklabel pos=right, grid style={thick,dashed,color=plotgray}},
			]
			\addplot[mark=none,color=plotblue] table[x=p,y=dlt] {super_era.csv};
		\end{axis}
	\end{tikzpicture}
	\caption{The quantity $\delta(\lm)$ defined in \eqref{eq:delta-lm}, as
		a function of $\lm$.
		Since $\delta(\lm)$ is a lower bound for 
		$\QC^{(1)}(\NC_{1/2}\otimes \EC_{\lm})-(\QC^{(1)}(\NC_{1/2})+\QC^{(1)}(\EC_{\lm}))$,
		the coherent information of $\NC_{1/2}$ and $\EC_{\lm}$ is (super)-amplified whenever
		$\delta(\lm)>0$, which holds when $\lm_{\mathrm{min}} \leq \lm \leq \lm_{\mathrm{max}}$.}   
	\label{fig:nonAddEra}
\end{figure}
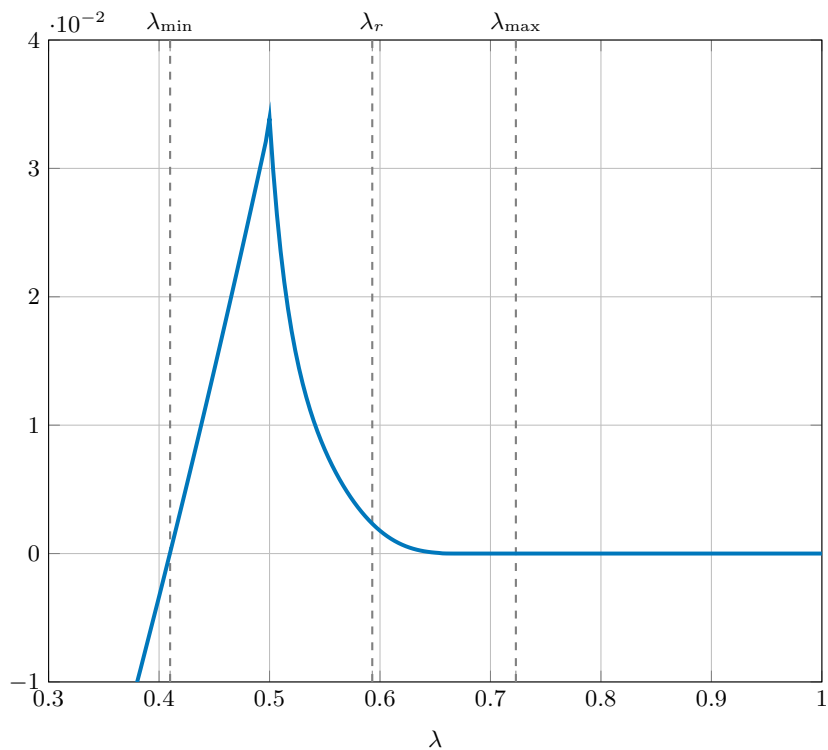

To demonstrate this (super-)amplification for $\NC_{1/2} \ot
\EC_{\lm}$, consider an input $\rho_{aa'}(\ep, r_1, r_2)$ of the form given in
\eqref{eq:ampNsAnz}, evaluate the coherent information on the input, and 
maximized over the parameters $\ep, r_1,$ and $r_2$:
\begin{align}
	\Dl^*(\NC_{1/2} \ot \EC_{\lm}) =  \max_{\ep, r_1, r_2} \Dl\big( \NC_{1/2} \ot
	\EC_{\lm} , \rho_{aa'}(\ep, r_1, r_2) \big).
	\label{eq:entMaxE}
\end{align}
Consider the quantity
\begin{align}
	\dl(\lm)\coloneqq \Dl^*(\NC_{1/2} \ot \EC_{\lm}) -  \left(\QC^{(1)}(\NC_{1/2}) + \QC^{(1)}(\EC_{\lm})\right)
	\label{eq:delta-lm}
\end{align}
whose strict positivity is a witness for amplification or superamplification of the
coherent information of $\NC_{1/2}$ and $\EC_{\lm}$ because
$\Dl^*(\NC_{1/2} \ot \EC_{\lm}) \leq \QC^{(1)}(\NC_{1/2} \ot \EC_{\lm})$.  
Our numerical optimization shows that $\dl(\lm)$ is positive when
$\lm_{\mathrm{min}} \leq \lm \leq \lm_r$
where $\lm_0 \simeq .41$ and $\lm_r \simeq .663$.  
The quantity $\dl(\lm)$ is increasing for $\lm_{\mathrm{min}} \leq \lm \leq 1/2$
and decreasing for $1/2 \leq \lm \leq \lm_r$, reaching a maximum value
of $\approx .033$ at $\lm = 1/2$.
In addition, a $\log$-singularity-based argument
\cite{Siddhu21} (see also Section \ref{sec:superNsElm}) 
shows that $\dl(\lm)>0$ for $\lm < \lm_{\mathrm{max}}$ where $\lm \simeq .723$.
Fig.~\ref{fig:nonAddEra} depicts $\delta(\lm)$ for all $\lm$ of
interest.  
In summary, we find amplification of coherent information of the joint channel $\NC_{1/2}\otimes \EC_{\lambda}$ for the large range $0.41 \lesssim \lambda \lesssim 0.72$ spanning both the degradable and antidegradable regimes of the erasure channel.  For $\lambda<1/2$, this finding demonstrates super-amplification of coherent information (complementary to the previous subsection).

Using the same arguments as before, the above holds also for quantum capacity subject to the spin-alignment
conjecture.  
This is a striking result: two channels $\NC_{1/2}$ and $\EC_{\lm}$,
each with \emph{non-zero} quantum capacity, together have {\em larger}
quantum capacity than the sum of each.

The mechanism behind this {\em super-amplification} of quantum
capacity in our work appears to be novel.
In the pioneering example of super-additivity of quantum
capacities~\cite{SmithYard08}, a (zero-capacity) symmetric channel is
employed to transform the private capacity of another channel into the
quantum capacity of the joint channel, at the cost of a factor of
$1/2$.
In particular, PPT channels with private capacity can be used to
establish \emph{p-bits} \cite{HorodeckiHorodeckiEA05,HorodeckiHorodeckiEA09}.
A p-bit is a unit of \emph{private} information and consists of a key and a shield system on each side.  The private
correlation resides in the key systems; the shield systems are 
correlated with both the key systems and the environment, in a way
that protects the key systems from the environment.
Furthermore, if the shield systems are brought to one of the two
parties, it can be decorrelated from the key systems, turning the key
into entanglement.
The protocol by Smith and Yard generates p-bits with a
PPT channel and transmits the sender's shield systems through the 50\%
erasure channel.  If the shield is transmitted, which happens half the
time, the p-bit can be transformed into entanglement.  If the shield is
erased, the coherent information generated is 0.
This is the origin of the factor of 1/2.
Amplification of quantum capacities is also demonstrated in
\cite{SmithSmolin09a,LeungLiEA14} using the same Smith-Yard
mechanism (note that an additional suppression of private
capacity and other tricks are used in \cite{SmithSmolin09a} to
attain an amplification bigger than half of the private capacity).

Even though the private capacity of $\NC_s$ is larger than its quantum capacity, we actually have $\QC(\NC_s) > \frac{1}{2}\PC(\NC_s)$.
It is therefore not advantageous to use the Smith-Yard protocol
because of the associated factor of 1/2.  We may expect that, in this
case, the best strategy to optimize the coherent information is to
put independent states into the erasure channel and $\NC_s$.  If this
strategy were optimal we would have additive coherent information
across $\NC_s$ and $\EC_{\lm}$.

Super-amplification of capacity here follows a more delicate
mechanism.
Examining the form of the input state \eqref{eq:ampNsAnz}, it appears to be
carrying out the Smith-Yard mechanism and the independent strategies
in superposition, while retaining a coherent memory of which strategy
was used.
The coherence between these two strategies is substantial,
providing additional coherent information that is otherwise
unattainable.
This exhibits the strengths of the two individual strategies without their corresponding drawbacks.

We also show similar results for the amplitude damping channel: the coherent information (and quantum capacity, subject to the SAC) of $\NC_s$ are (super-)amplified by the amplitude damping channel $\AC_\gamma$ with similar magnitude and over a similar range of the noise parameter $\gamma$.
Since this is again achieved by the same input
\eqref{eq:ampNsAnz}, the (super-)amplification is caused by the same mechanism as detailed above.

\subsection{(Super)-amplification of coherent information of $\NC_s$}
\label{sec:superNs}

In this subsection, we study the (super-)amplification of coherent
information when $\NC_s$ is used jointly with an additional
channel $\aux_x$, which is either 
an erasure channel $\EC_\lambda$,
an amplitude damping channel $\AC_\gamma$,
or a depolarizing channel $\DC_p$. 
We let both parameters $s$ and $x$ vary and chart out the
two-dimensional regions $(s,x)$ in which (super-)amplification of
coherent information can be demonstrated.

As an overview, our analysis will be similar to those presented in
Section \ref{sec:erasure}, except now the two channels of interest are
$\NC_s$ and $\aux_x$.  We fix an arbitrary $s$ and find the range of
$x$ for which we can verify (super)-amplification, using both
numerical methods and $\log$-singularity based arguments.  The
verification relies on the positivity of a quantity $\dl(\NC_s, \aux_x)$ (defined below in \eqref{eq:gen-diff}) that generalizes
$\delta(\lm)$ in \eqref{eq:delta-lm}.  
This generalized quantity depends on both parameters $s$ and $x$; as a function of
$x$ for fixed $s$, it behaves similarly to $\delta(\lm)$ as depicted in
Fig.~\ref{fig:nonAddEra}.

In more detail, recall that $\HC_a$ and $\HC_{a'}$ denote the input
spaces of $\NC_s$ and $\aux_x$ respectively.  Consider again the density
operator $\rho_{aa'}(\ep, r_1, r_2)$ given in \eqref{eq:ampNsAnz}, 
and the coherent information of $\NC_s \ot \aux_x$ optimized over the parameters
of $\rho_{aa'}$:
\begin{align}
	\Dl^*(\NC_s \ot \aux_x) = \underset{\ep, r_1, r_2}{\max} \Dl\big( \NC_s \ot \aux_x, \rho_{aa'}(\ep, r_1, r_2) \big).
	\label{eq:gen-Max-del}
\end{align}
We can generalize $\delta(\lm)$ in \eqref{eq:delta-lm} to the following
quantity:
\begin{align}
	\dl(\NC_s, \aux_x) = 
	\Dl^*(\NC_s \ot \aux_x) - \left( \QC^{(1)}(\NC_s) + \QC^{(1)}(\aux_x) \right) \,. 
	\label{eq:gen-diff}
\end{align}
As before, since
$\Dl^*(\NC_s \ot \aux_x) \leq \QC^{(1)}(\NC_s \ot \aux_x)$, the
positivity of $\dl(\NC_s, \aux_x)$ verifies the (super)-amplification
of coherent information of $\NC_s$ and $\aux_x$.  For each $\aux_x$,
the quantity $\dl(\NC_s, \aux_x)$ is found to be positive for an interval of $x$ denoted as $[x_{\mathrm{min}}(s),x_{\mathrm{max}}(s)]$.  
Within this range, $\dl(\NC_s, \aux_x)$ increases with $x$ reaching a substantial maximum
and vanishes numerically, and a $\log$-singularity based argument is
applied to verify positivity until $x=x_{\mathrm{max}}(s)$.  Repeating this
analysis for every $s$ gives the region $(s,x)$ for
(super)-amplification.  We are now ready to apply this framework to
the three families of additional channel of interest.

\subsubsection{(Super)-amplification with erasure channel}
\label{sec:superNsElm}

We numerically find that the threshold noise rate $\lm_{\mathrm{min}}(s) < 1/2$ for
all $s \in (0,1/2]$.
Therefore, the erasure channel \emph{super-amplifies}
the coherent information for all such $s$.  The maximum of
$\dl(\NC_s, \EC_{\lm})$ is attained at $\lm = 1/2$.  The numerical
value of $\dl(\NC_s, \EC_{\lm})$ vanishes at some $\lm>1/2$, and so we switch to a method based on $\log$-singularity~\cite{Siddhu21} to obtain $\lm_{\mathrm{max}}(s)$.

We now review the ideas in \cite{Siddhu21}.  Let $\sigma(\ep)$ denote
a density operator that depends on a real parameter $\ep$.  If some
eigenvalues of $\sigma(\ep)$ increase linearly from zero to leading
order in $\ep$, the von Neumann entropy, $S(\sigma(\ep)) =
- \Tr \big( \sigma(\ep) \log \sigma(\ep) \big)$, increases by an
amount $\simeq r |\ep \log \ep|$ to leading order in $\ep$ for some
$r>0$.  This increase is discontinuous at $\ep=0$; for small $\ep$, we
call the increase a $\log$-singularity, and the value $r$ is called the
{\em rate} of the singularity.  We also say that $S(\sigma(\ep))$ has
an $\ep$-$\log$-singularity of rate $r$.  These ideas are easily generalized to linear functions of entropies.

We now apply the above ideas on $\log$-singularity to show 
$\dl(\NC_s, \EC_{\lm})>0$ for certain values of $\lm$.  
First note that
$\rho_{aa'}(0,r_1,0) = \left( r_1 [0]+ (1-r_1) [2] \right)_a \ot [0]_{a'}$. 
Following \eqref{eq:NsQ1} there is a choice of $r_1$ (call it $u$)
so that $r_1 [0]+ (1-r_1) [2]$ attains the coherent information
of $\NC_s$.  Furthermore, for $\lm \geq 1/2$, following
\eqref{eq:ErQ1Q}, $[0]$ attains the coherent information of
$\EC_{\lm}$.  Therefore, 
$\Dl(\NC_s \ot \EC_{\lm}, \rho_{aa'}(0,u,0))
= \QC^{(1)}(\NC_s) + \QC^{(1)}(\EC_{\lm})$ and the difference 
\begin{align}
	\Dl(\NC_s \ot \EC_{\lm}, \rho_{aa'}(\ep,u,0)) - \Dl(\NC_s \ot \EC_{\lm}, \rho_{aa'}(0,u,0))
	\label{eq:del-erasure-lb}
\end{align}
is a lower bound on $\delta(\NC_s,\EC_{\lm})$.  
Note that 
\begin{align}
	\Dl(\NC_s \ot \EC_{\lm}, \rho_{aa'}(\ep,u,0)) =
	S(\rho_{bb'}(\ep,u,0)) - S(\rho_{cc'}(\ep,u,0)) 
\end{align}
where 
\begin{align}
	\rho_{bb'}(\ep,u,0) &= (\NC_s \ot \EC_{\lm})(\rho_{aa'}(\ep,u,0)), & 
	\rho_{cc'}(\ep,u,0) &= (\NC_s^c \ot \EC_{\lm}^c)(\rho_{aa'}(\ep,u,0)).
\end{align}
Using a suitable basis of $\HC_{bb'}$ and $\HC_{cc'}$, one can show that
\begin{align}
	\begin{aligned}
		\rho_{bb'}(\ep,u,0) &= {\rm diag} \big(s(1-u)(1-\lm), 0, s(1-u)\lm \big) \\
		&\qquad \oplus {\rm diag} \big((1-s)(1-u)(1-\lm),0, (1-s)(1-u)\lm \big) \\
		&\qquad \oplus {\rm diag} \big((1-\ep) u (1-\lm), \ep u (1-\lm), u\lm \big)\\
		\rho_{cc'}(\ep,u,0) &= {\rm diag} \big(s(1-u)\lm, (s(1-u) + u
		\ep)(1-\lm), \big) \\
		&\qquad \oplus {\rm diag} \big(0, (1 - s(1-u) -u \ep)(1-\lm) \big) \\
		&\qquad \oplus 
		\begin{pmatrix}
			\ep u \lm & u \lm \sqrt{\ep (1-\ep)} \\
			u \lm \sqrt{\ep (1-\ep)} & (1 - s(1-u) -u\ep)\lm
		\end{pmatrix},
	\end{aligned}
\end{align}
where ${\rm diag} (d_1, d_2, \cdots)$ represents a diagonal square matrix of
diagonal entries $d_1, d_2, \cdots$, and the operation $\oplus$
performs the direct sum of matrices.  
From the above expressions, one can conclude that
$S(\rho_{bb'}(\ep,u,0))$ has an $\ep$-$\log$-singularity of rate
$u(1-\lm)$ and $S(\rho_{cc'}(\ep,u,0))$ has an $\ep$-$\log$-singularity of rate $u \lm (1-u)(1-s)/\big( 1- s(1-u)\big)$.
The coherent information $\Dl(\NC_s \ot \EC_{\lm}, \rho_{aa'}(\ep,u,0))$
thus has an $\ep$-$\log$-singularity of positive rate if
\begin{align}
	u(1-\lm) &> u \lm (1-u)(1-s)/\big( 1- s(1-u)\big)\\
	\intertext{or equivalently,}
	\lm &< \frac{1 - s + us}{2-2s-u+2us} \eqqcolon  \lm_{\mathrm{max}}(s), 
\end{align}
in which case \eqref{eq:del-erasure-lb} is strictly positive
for small $\ep>0$.  This implies $\dl(\NC_s,\EC_{\lm})>0$ for 
$0 < s \leq 1/2$ and $1/2 \leq \lm < \lm_{\mathrm{max}}(s)$, demonstrating
\emph{amplification} of coherent information of $(\NC_s,\EC_{\lm})$. 

We summarize our findings in this subsection by plotting $\lm_{\mathrm{min}}(s)$
and $\lm_{\mathrm{max}}(s)$ as a function of $s$ in Fig.~\ref{fig:lm0lm1};
each point $(s,\lm)$ between these two curves in the orange-shaded region corresponds to
a pair of channels $\NC_s, \EC_\lm$ with provable 
(super)-amplification of coherent information.

\begin{figure}[ht]
	\centering
	\begin{tikzpicture}
		\begin{axis}[
			scale = 1.4,
			xmin = 0,
			xmax =0.5,
			xlabel = $s$,
			ylabel = $\lambda$,
			ylabel style={rotate=-90},
			every axis plot/.append style={line width=1.25pt},
			x tick label style={/pgf/number format/fixed},
			grid = both,
			]
			\addplot[dashed,color=plotgray,line width = 1pt] {0.5};
			\addplot[name path = lower,mark=none,color=plotblue] table[col sep = comma,x=s,y=lmMin] {NsElNonAddRegion_1.csv} node[pos=0.2,below=1em,font=\large] {$\lambda_{\mathrm{min}}$};
			\addplot[name path = upper,mark=none,color=plotmagenta] table[col sep = comma,x=s,y=lmMax] {NsElNonAddRegion_1.csv} node[pos=0.2,above=.7em,font=\large] {$\lambda_{\mathrm{max}}$};
			\addplot[plotorange!50,fill opacity = 0.4] fill between [of = lower and upper];
		\end{axis}
		\begin{axis}[
			xshift = 5.9cm,
			yshift = 3.5cm,
			width = 4.75cm,
			xmin = 0,
			xmax = 0.3,
			every axis plot/.append style={line width=1.25pt},
			axis background/.style={fill=white},
			y tick label style={/pgf/number format/.cd,precision=3},
			tick label style={font=\small},
			grid = both,
			]
			\addplot[dashed,color=plotgray,line width = 1pt] {0.5};
			\addplot[name path = lower,mark=none,color=plotblue] table[col sep = comma,x=s,y=lmMin] {NsElNonAddRegion_1.csv} node[pos=0.1,below] {$\lambda_{\mathrm{min}}$};
		\end{axis}
	\end{tikzpicture}
	\caption{Region in the $\lambda$-$s$ plane where $\QC^{(1)}(\NC_s \ot \EC_{\lm})$ is
		super-additive (shaded in orange). 
		The boundaries of the shaded region are the minimal ($\lambda_{\mathrm{min}}$, blue) and maximal ($\lambda_{\mathrm{max}}$, magenta) values of the erasure probability $\lambda$ for which super-additivity occurs for fixed $s$.
		The inset plot zooms into the interval $s\in[0,0.3]$, showing that $\lambda_{\mathrm{min}} < 1/2$ for all $s\in[0,0.5]$.}
	\label{fig:lm0lm1}
\end{figure}

\subsubsection{(Super)-amplification with amplitude damping channel}
\label{sec:superAd}

We repeat the analysis of the previous subsection for the amplitude dampling channel $\AC_\gamma$
instead of the erasure channel, and obtain very similar results.  
For any fixed $s$ in $(0,1/2]$, the quantity $\dl(\NC_s, \AC_\gamma)$ is positive
for $\gamma_{\mathrm{min}}(s) \leq \gamma \leq \gamma_{\mathrm{max}}(s)$ (see
Fig.~\ref{fig:p0p1}).  Since $\gamma_{\mathrm{min}}(s) < 1/2$, the coherent
information of any $\NC_s$ can be \emph{super-amplified} by some
amplitude damping channel (using similar arguments as before).  The
numerical value of $\dl(\NC_s, \AC_\gamma)$ first increases with
$\gamma$, reaches a maximum, and then decreases and vanishes.  A
$\log$-singularity based argument provides the value of
$\gamma_{\mathrm{max}}(s)$~\cite{Siddhu21},
\begin{align}
	\gamma_{\mathrm{max}}(s) &= \frac{1}{1 + k}\qquad \text{with}\quad k = \frac{(1-s)(1-u)}{u + (1-s)(1-u)},\label{eq:p1s}
\end{align}
and $u$ is chosen so that $(1-u)[0] + u[1]$ attains the
coherent information of $\AC_\gamma$ (see \eqref{eq:q1Acp}).  
We have $\gamma_{\mathrm{max}}(s) \geq 1/2$ for all $s \in (0,1/2]$, and hence the channel $\AC_\gamma$ \emph{amplifies} the coherent information of $\NC_s$ for $1/2 \leq \gamma \leq \gamma_{\mathrm{max}}(s)$.

\begin{figure}[ht]
	\centering
	\begin{tikzpicture}
		\begin{axis}[
			scale = 1.4,
			xmin = 0,
			xmax =0.5,
			xlabel = $s$,
			ylabel = $\gamma$,
			ylabel style={rotate=-90},
			every axis plot/.append style={line width=1.25pt},
			x tick label style={/pgf/number format/fixed},
			grid = both,
			]
			\addplot[dashed,color=plotgray,line width = 1pt] {0.5};
			\addplot[name path = lower,mark=none,color=plotblue] table[x=s,y=pMin] {NsApNonAddRegion_2.csv} node[pos=0.1,below=1em,font=\large] {$\gamma_{\mathrm{min}}$};
			\addplot[name path = upper,mark=none,color=plotmagenta] table[x=s,y=pMax] {NsApNonAddRegion_2.csv} node[pos=0.1,above=.7em,font=\large] {$\gamma_{\mathrm{max}}$};
			\addplot[plotorange!50,fill opacity = 0.4] fill between [of = lower and upper];
		\end{axis}
		\begin{axis}[
			xshift = 5.85cm,
			yshift = 3.7cm,
			width = 4.75cm,
			xmin = 0,
			xmax = 0.2,
			every axis plot/.append style={line width=1.25pt},
			axis background/.style={fill=white},
			y tick label style={/pgf/number format/.cd,precision=4},
			x tick label style={/pgf/number format/fixed},
			tick label style={font=\small},
			grid = both,
			]
			\addplot[dashed,color=plotgray,line width = 1pt] {0.5};
			\addplot[name path = lower,mark=none,color=plotblue] table[x=s,y=pMin] {NsApNonAddRegion_2.csv} node[pos=0.1,below] {$\gamma_{\mathrm{min}}$};
		\end{axis}
	\end{tikzpicture}
	\caption{Region in the $\gamma$-$s$ plane where $\QC^{(1)}(\NC_s \ot \AC_{\gamma})$ is
		super-additive (shaded in orange). 
		The boundaries of the shaded region are the minimal ($\gamma_{\mathrm{min}}$, blue) and maximal ($\gamma_{\mathrm{max}}$, magenta) values of the amplitude damping probability $\gamma$ for which super-additivity occurs for fixed $s$.
		The inset plot zooms into the interval $s\in[0,0.2]$, showing that $\gamma_{\mathrm{min}} < 1/2$ for all $s\in[0,0.5]$.}
	\label{fig:p0p1}
\end{figure}

\subsubsection{(Super)-amplification with depolarizing channel}
\label{sec:superAdDp}

\begin{figure}[ht]
	\centering
	\begin{tikzpicture}
		\begin{axis}[
			xlabel=$s$,
			xmin = 0.44,
			xmax = 0.5,
			ylabel=$p$,
			ylabel style={rotate=-90},
			ymin = 0.14,
			ymax = 0.21,
			scale=1.25,
			every axis plot/.append style={line width=1.5pt},
			grid = both,
			]
			\addplot[color=plotgray,line width=1pt, dashed, domain=0.44:0.5] {.18928962492} node[below,pos = 0.1,font=\large,color=plotgray] {$p^*$};
			\addplot[name path = lower,mark=none,color=plotblue] table[x=s,y=pMin] {NsDpNonAddRegion.csv} node[below=1em,pos = 0.2,font=\large] {$p_{\mathrm{min}}$};
			\addplot[name path = upper,mark=none,color=plotmagenta] table[x=s,y=pMax] {NsDpNonAddRegion.csv} node[above=.7em,pos = 0.2,font=\large] {$p_{\mathrm{max}}$};
			\addplot[plotorange!50,fill opacity = 0.4] fill between [of = lower and upper];
		\end{axis}
	\end{tikzpicture}
	\caption{
		Region in the $p\,$-$s$ plane where $\QC^{(1)}(\NC_s \ot \DC_{p})$ is super-additive (shaded in orange).
		The boundaries of the shaded region are the minimal ($p_{\mathrm{min}}$, blue) and maximal ($p_{\mathrm{max}}$, magenta) values of the depolarizing probability $p$ for which super-additivity occurs for fixed $s$.
		In contrast to the erasure channel and amplitude damping channel above, super-additivity is limited to the interval $s\in[0.4539,1/2]$.
		The dashed line marks the hashing point $p^*\simeq 0.1893$ at which $\QC^{(1)}(\DC_{p^*}) = 0$.}
	\label{fig:Dp0p1}
\end{figure}
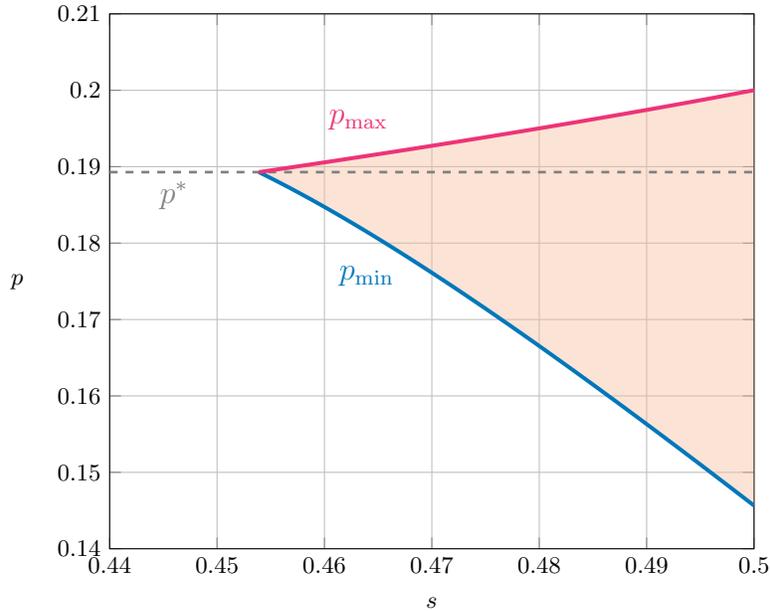

Finally, we study (super)-amplification of coherent information
between $\NC_s$ and the depolarizing channel $\DC_p$.  Numerically,
$\dl(\NC_s, \DC_{p})$ is positive for some $p$ only if $s \in [s_{\rm
	min},1/2]$, where $s_{\rm min} \simeq .4539$.  For each $s$ in this
interval, $\dl(\NC_s, \DC_{p})>0$ for $p_{\mathrm{min}}(s) \leq p \leq p_{\mathrm{max}}(s)$
(see Fig.~\ref{fig:Dp0p1}).  Since $p_{\mathrm{min}}(s) \leq p^* \leq p_{\mathrm{max}}(s)$, where $p^* \simeq 0.1893$ is the smallest $p$ for
$\QC^{(1)}(\DC_p) = 0$, both super-amplification and amplification of
coherent information are demonstrated for each $s \in [s_{\rm
	min},1/2]$.  We note that $\dl(\NC_s, \DC_{p})$ is increasing
for $p_{\mathrm{min}}(s) \leq p \leq p^*$ and decreasing for
$p^* \leq p \leq p_{\mathrm{max}}(s)$, reaching a maximum at $p = p^*$.

\subsection{Universal amplification}
\label{sec:randchan}

In the sections above we showed how one single coding strategy (the input state in \eqref{eq:ampNsAnz}) achieves amplification of quantum capacity of $\NC_{s}$ when used in combination with an amplitude damping, erasure, or depolarizing channel.
However, our results raise the following rather natural question: to what extent is the amplification mechanism making use of the symmetries of the channel used in conjunction with $\NC_{s}$?
In particular, both the depolarizing and erasure channel have full unitary covariance at the input; an amplification code ansatz tailored to the amplitude damping channel could thus make use of this covariance property and ``coincidentally'' enable the amplification effect for erasure and depolarizing noise.
In this section, we show that this is \emph{not} the case: we provide
numerical results demonstrating that the code
ansatz \eqref{eq:ampNsAnz} successfully amplifies the
coherent information of $\NC_{1/2}$ when used together with a \emph{random} qubit channel.

Before explaining this result in more detail, we briefly summarize
the linear representation of qubit-qubit channels based on the
``Bloch vector'' as used in, e.g., \cite{KingRuskai01}.
This representation of quantum channels uses the Bloch vector
representation for qubit states: a qubit state $\rho$ is written as $\rho = \frac{1}{2}(I + s_x X + s_y Y + s_z Z)$, where $\mathbf{s} = (s_x,s_y,s_z)\in\mathbb{R}^3$ is the Bloch vector with $\|\mathbf{s}\|\leq 1$.
The state $\rho$ is pure if and only if $\|\mathbf{s}\|=1$.
With the identification $\rho \longleftrightarrow r(\rho) = \frac{1}{2}(1,\mathbf{s})$, any linear map $\FC\colon M_2(\mathbb{C})\to M_2(\mathbb{C})$
from the space $M_2(\mathbb{C})$ of complex $(2\times2)$-matrices into itself can be represented as a $(4\times 4)$-matrix $F\in M_4(\mathbb{C})$ acting on the vector representation $r(\rho)$ of $\rho$.
Assuming that $\FC$ is a quantum channel and hence trace-preserving and completely positive, the linear representation $F$ has the following form:
\begin{align}
	F = \begin{pmatrix}	
		1 & \mathbf{0}\\
		\mathbf{t} & T
	\end{pmatrix},\label{eq:channel-bloch-rep}
\end{align}
where $\mathbf{t}\in\mathbb{R}^3$ and $T\in M_3(\mathbb{R})$ is a matrix whose eigenvalues satisfy a certain algebraic condition ensuring complete positivity of $\FC$ \cite{KingRuskai01}.
Every qubit-qubit quantum channel $\FC\colon M_2(\mathbb{C})\to M_2(\mathbb{C})$
is unitarily equivalent to a channel $\RC$ with matrix representation \eqref{eq:channel-bloch-rep} in ``normal form''
\begin{align}
	R = \begin{pmatrix}
		1 & 0 & 0 & 0\\
		t_1 & \lambda_1 & 0 & 0\\
		t_2 & 0 & \lambda_2 & 0\\
		t_3 & 0 & 0 & \lambda_3
	\end{pmatrix};\label{eq:channel-normal-form}
\end{align}
again, the vectors $\mathbf{t}=(t_1,t_2,t_3)$ and $\lambda=(\lambda_1,\lambda_2,\lambda_3)$ need to satisfy certain conditions for $\RC$ to be completely positive.
Note that a channel is unital if and only if $\mathbf{t}=\mathbf{0}$ in \eqref{eq:channel-bloch-rep}; in this case, a channel in normal form \eqref{eq:channel-normal-form} corresponds to a Pauli channel, i.e., $\RC(\rho) = p_0 \rho + p_1 X\rho X + p_2 Y\rho Y + p_3 Z\rho Z$ for some probability distribution $(p_0,p_1,p_2,p_3)$, with each $p_i$ a function of $\lambda$ \cite{KingRuskai01}.

Having introduced the representation \eqref{eq:channel-bloch-rep} of qubit-qubit channels and their normal form \eqref{eq:channel-normal-form}, we can now outline our numerical procedure to demonstrate the universal character of the amplification strategy \eqref{eq:ampNsAnz} for $\NC_{1/2}$:
\begin{enumerate}
	\item Randomly sample $\mathbf{t}=(t_1,t_2,t_3)$ and $\lambda = (\lambda_1,\lambda_2,\lambda_3)$, ensuring that the map $\RC$ defined through its matrix $R$ as in \eqref{eq:channel-normal-form} is completely positive and hence a quantum channel.
	\item Form the 1-parameter family of channels $\RC_x = (1-x) \id + x \RC$, where $x\in [0,1]$. 
	\item Choose $x^*$ to be the ``hashing point'' of $\RC_x$, i.e., the smallest $x\in [0,1]$ such that $\QC^{(1)}(\RC_x) = 0$; this is easily achieved as $x\mapsto \QC^{(1)}(\RC_x)$ is typically monotonically decreasing.
	\item Evaluate the coherent information $\Dl(\NC_{1/2} \otimes \RC_{x^*},\rho)$, where $\rho$ is the amplification code ansatz from \eqref{eq:ampNsAnz}, and test whether 
	\begin{align}
		\QC^{(1)}(\NC_{1/2} \otimes \RC_{x^*}) \geq \Dl(\NC_{1/2} \otimes \RC_{x^*},\rho) > \QC^{(1)}(\NC_{1/2}) = \QC^{(1)}(\NC_{1/2}) + \QC^{(1)}(\RC_{x^*}).\label{eq:amplification}
	\end{align}
\end{enumerate}
We find that in a substantial fraction of test runs, the above procedure produces a qubit-qubit channel $\RC_{x^*}$ for which the strict inequality in \eqref{eq:amplification} indeed holds.
Super-additivity typically persists in a non-trivial interval around the hashing point $x^*$.
In particular, we also find super-additivity of coherent information for $\NC_{1/2}\otimes \RC_x$ for some $x<x^*$ such that $\QC^{(1)}(\RC_x) > 0$.
The channel $\RC$ used in the definition of $\RC_{x}$ is randomly sampled, and hence has no special structure other than the normal form \eqref{eq:channel-normal-form}.
This demonstrates the universal character of our amplification mechanism (facilitated by the code ansatz \eqref{eq:ampNsAnz} and explained in Sec.~\ref{sec:erasure}) in a clean and compelling way.

As a concrete example, consider the channel $\RC$ with corresponding matrix representation 
\begin{align}
	R = \begin{pmatrix}
		1 & 0 & 0 & 0\\
		0.0078 & 0.1253 & 0 & 0\\
		0.4231 & 0 & 0.1302 & 0\\
		0.6556 & 0 & 0 & 0.0924
	\end{pmatrix}.\label{eq:concrete-channel}
\end{align}
The hashing point of the channel family $\RC_x = (1-x)\id + x \RC$ lies in the interval $[0.3061,0.3265]$, with $\QC^{(1)}(\RC_{x^*}) = 0$ for $x^*= 0.3265$.
At this noise level we achieve the following coherent information using the ansatz \eqref{eq:ampNsAnz} for the input state $\rho$:
\begin{align}
	\Dl(\NC_{1/2} \otimes \RC_{x^*},\rho) = 0.7175 > 0.6942 = \QC^{(1)}(\NC_{1/2}) + \QC^{(1)}(\RC_{x^*}).
\end{align}
This super-additivity of coherent information achieved by the ansatz \eqref{eq:ampNsAnz} persists in the interval $x\in[0.246,0.365]$.

The normal form \eqref{eq:channel-normal-form} is not restrictive and chosen simply for convenience;
the amplification of coherent information can also be observed when randomly sampling a channel of the more general form \eqref{eq:channel-bloch-rep}.
Finally, we note that the amplification of coherent information can typically be enhanced further when searching over all possible codes for the joint channel $\NC_{1/2} \otimes \RC_{x^*}$; this shows that, while our amplification mechanism \eqref{eq:ampNsAnz} is in a certain sense universal, it is by no means optimal in general.
For example, for the concrete channel $\RC_x$ defined using \eqref{eq:concrete-channel}, an even higher coherent information of $\Dl(\NC_{1/2} \otimes \RC_{x^*},\tau) = 0.7241$ is achieved by the input state $\tau_{aa'} = p [0]_a \otimes \sigma_{a'} + (1-p) [\psi]_{aa'}$ with $p = 0.5281$ and
\begin{align}
	\begin{aligned}
		\sigma_{a'} &= \begin{pmatrix}
			0.7434 & 0.0032 - 0.1702i\\  0.0032 + 0.1702i & 0.2566
		\end{pmatrix}
		\\
		|\psi\rangle_{aa'} &= (0.4805+0.1806i) |1\rangle_a |0\rangle_{a'} + (0.3184-0.1260i) |1\rangle_{a} |1\rangle_{a'}\\
		& \quad {} + (-0.2938-0.5865i) |2\rangle_{a} |0\rangle_{a'} + 0.4348 |2\rangle_{a} |1\rangle_{a'}.
	\end{aligned}
\end{align}
For this example channel $\RC_x = (1-x)\id + x \RC$ with $\RC$ as in \eqref{eq:concrete-channel}, optimizing over all codes extends the range of super-additivity of coherent information to $x\in[0.228,0.383]$.
For details of these computations, see the code repository \cite{github} for this paper.

\section{Violation of strong additivity of quantum capacity involving $\MC_d$}
\label{sec:amplification-Md}

In the last section, we studied the coherent information and the quantum
capacity of $\NC_s$ when used in parallel with other
channels.
In this section, we carry out a similar analysis for another family $\MC_{d}$
of channels generalizing $\NC_{1/2}$ (fixing $s=1/2$) to arbitrary input dimension $d$.
For the additional channel, we focus on $\EC_{\lm,d}$, the
$d$-dimensional generalization of the erasure channel with erasure
probability $\lm$.  
We obtain the following results:
\begin{enumerate}[(1)]
	\item Similar to the analysis for $\NC_s$, 
	we yet again find super-additivity of the channel coherent
	information for all $d \geq 3$ and erasure probability $\lm=1/2$.
	Assuming the validity of the spin-alignment conjecture, this
	super-additivity can be elevated to the level of quantum capacity.
	\item An additional feature offered by $\MC_{d}$ is a nice upper bound for the quantum capacity of $\MC_{d}$ as a (decreasing) function of $d$, proved in the companion paper \cite{paper1IEEE}.
	This bound allows superadditivity of quantum capacity to be proved \emph{unconditionally} (without the need for the spin alignment conjecture), and for the \emph{full range} of $\lambda$! 
	More specifically, superadditivity of quantum capacity can be proved for:   
	\begin{enumerate}[(a)]
		\item erasure probability $\lambda \in [0,37,0.57]$ and moderate values
		of $d$ (depending on $\lambda$);
		\item any $0<\lambda<1$ for sufficiently large $d$ using a $\log$-singularity argument. 
	\end{enumerate}
\end{enumerate}

\subsection{The $\MC_d$ channel and the qudit erasure channel $\EC_{\lm,d}$}
\label{sec:MdChan}

The isometry $F_s$ in~\eqref{isoDef1} with $s=1/2$ has a higher-dimensional generalization $G$ taking $\HC_a$ to $\HC_b \ot \HC_c$,
where $\HC_a$, $\HC_b$ and $\HC_c$ have dimensions $d$, $d$, and $d-1$,
respectively.  This generalization acts as
\begin{align}
	G \ket{0} = \frac{1}{\sqrt{d-1}}\sum_{j=0}^{d-2} \ket{j} \ot \ket{j}, \quad 
	G \ket{i} = \ket{d-1} \ot \ket{i-1},
	\label{isoDef2}
\end{align}
for $1 \leq i \leq d-1$.  For $d=3$ this is the isometry from Sec.~\ref{sec:NsChan}, i.e., $G = F_{1/2}$.  When 
$d \geq 3$, the isometry $G$ generates a pair of channels $(\MC_d, \MC_d^c)$
generalizing the pair $(\NC_{1/2},\NC_{1/2}^c)$.  
We prove in Sections 3.2 and 8.2 of \cite{paper1IEEE} that 
$\QC(\MC_d^c)  = \PC(\MC_d^c) = \CC(\MC_d^c) = \log (d-1)$ and $\PC(\MC_d) = \CC(\MC_d) = 1$.  

We mostly focus on the channel coherent information and the quantum capacity
here.  In Section 4 of~\cite{paper1IEEE}, we show that for any $d \geq 3$,
\begin{align}
	\QC^{(1)}(\MC_d) = \max_{\rho_a(u)} \Dl\big( \MC_d, \rho_a(u) \big),
	\label{eq:MdQ1}
\end{align}
where $\rho_a(u) = (1-u)[0] + u [i]$, $0 \leq u \leq 1$ and $i$ is any fixed
integer between $1$ and $d-1$.
The channel coherent information described above can be further evaluated:  
\begin{align}
	\QC^{(1)}(\MC_{d}) &= \max_{0 \leq u \leq 1} \big( h(u) + (1-u) \log (d-1) +
	g(u,d-1) \big),
	\label{eq:q1Md}
	\intertext{where $h(\cdot)$ is the binary entropy function, and} 
	g(u,d) &= (d-1) \; \eta \! \left(\frac{1-u}{d}\right) + \eta \! \left(\frac{1-u}{d} + u \right), 
	\label{eq:q1MdSu}
\end{align}
with $\eta(x)\coloneqq x \log x$.

To see this, we first record the following useful identity, which is valid for $x\in[0,1]$ and any two states $\rho,\sigma$ with orthogonal support:
\begin{align}
	S((1-x)\rho + x \sigma) = h(x) + (1-x)S(\rho) + x S(\sigma),\label{eq:entropy-orthogonal}
\end{align}
Setting $\rho_a(u) = (1-u)[0] + u [1]$, we use \eqref{isoDef2} to compute the action of the channel $\MC_d$ on the input state $\rho_a(u)$:
\begin{align}
	\MC_d(\rho_a(u)) &= (1-u) \frac{1}{d-1} \sum_{j=0}^{d-2}[j]_b + u [d-1]_b,
	\intertext{which using \eqref{eq:entropy-orthogonal} has entropy}
	S(\MC_d(\rho_a(u))) &= h(u) + (1-u) \log(d-1).\label{eq:ent-rhob}
\end{align}
On the other hand, the complementary channel $\MC_d^c$ acting on $\rho_a(u)$ yields a state
\begin{align}
	\MC_d^c(\rho_a(u)) &= (1-u)\frac{1}{d-1} \sum_{j=0}^{d-2} [j]_c + u[0]_c = \left(\frac{1-u}{d-1} + u\right) [0]_c + \frac{1-u}{d-1}\sum_{j=1}^{d-2} [j]_c
	\intertext{with entropy}
	S(\MC_d^c(\rho_a(u))) &= - \eta\left(\frac{1-u}{d-1}+u\right) - (d-2) \eta\left(\frac{1-u}{d-1}\right).\label{eq:ent-rhoc}
\end{align}
The difference of \eqref{eq:ent-rhob} and \eqref{eq:ent-rhoc} yields the expression for the channel coherent information in \eqref{eq:q1Md}.

Subject to the spin-aligment conjecture in~\cite{paper1IEEE},
\begin{align}
	\QC(\MC_d) = \QC^{(1)}(\MC_d).
\end{align}
Both $\NC_{1/2}$ and $\MC_d$ have non-zero quantum capacity~(see Sec.~4
of~\cite{paper1IEEE}).
We also have an upper bound of the quantum capacity from Section 7.2 of
\cite{paper1IEEE}:  
\begin{align}
	\QC(\MC_d) \leq \log\left(1+\frac{1}{\sqrt{d-1}}\right)
	\leq {1 \over \ln 2} \frac{1}{\sqrt{d-1}} \,.\label{eq:Md-Q-bound}
\end{align}

In the following we consider the $d$-dimensional generalization of the erasure channel,
$\EC_{\lm,d}$, which replaces a $d$-dimensional input by an erasure symbol
$[e]$ (orthogonal to the input space) with probability $\lm$ and transmits the
input noiselessly with probability $1-\lm$.  
Similar to the qubit case, $\EC_{\lm,d}^c = \EC_{1-\lm,d}$, and $\EC_{\lm,d}$
is degradable for $0 \leq \lm \leq 1/2$ and anti-degradable for $1/2 \leq \lm
\leq 1$. The quantum capacity of $\EC_{\lm,d}$ is given by an analogue
of~\eqref{eq:ErQ1Q} \cite{BennettDiVincenzoEA97},
\begin{align}
	\QC^{(1)}(\EC_{\lm,d}) = \QC(\EC_{\lm,d}) = \max\lbrace (1-2\lambda)\log d,0\rbrace.
	\label{eq:ErQ1Qd}
\end{align}

\subsection{Amplification of $\MC_{d{+}1}$ with the qudit erasure channel $\EC_{\lm,d}$}

Let $\HC_r$, $\HC_{r'}$, $\HC_a$ and $\HC_{a'}$ be Hilbert spaces of dimension $2$, $d$, $d+1$, and $d$, respectively, with $\HC_a$ and $\HC_{a'}$ being the input spaces for $\MC_{d+1}$ and $\EC_{\lm,d}$, respectively. 
Consider a pure state $\ket{\psi} \in \HC_a \otimes \HC_r \otimes \HC_{a'}
\otimes \HC_{r'}$ of the following form:
\begin{align}
   \ket{\psi}_{ara'r'} = \sqrt{1-w} \; \ket{0}_r\ket{0}_a
    \; \frac{1}{\sqrt{d}} \sum_{i=0}^{d-1} \ket{i}_{r'} \ket{i}_{a'}
	+ 
    \sqrt{w} \; \ket{1}_{r}\ket{0}_{r'} \; \frac{1}{\sqrt{d}} \sum_{i=1}^d \ket{i}_{a} \ket{i-1}_{a'} .\label{eq:pure-input-state}
\end{align}
Taking the partial trace of the projector $[\psi]_{ara'r'}$ over $\HC_r$ and $\HC_{r'}$ results in a density operator $\rho_{aa'}$
on the input space of $\MC_{d+1} \ot \EC_{\lm}^d$.  

We now show that the coherent information $\Dl(\MC_{d+1} \ot \EC_{\lm,d}, \rho_{aa'})$ is given by the following expression:
\begin{align}
	\Dl(\MC_{d+1} \ot \EC_{\lm,d}, \rho_{aa'}) &= 
	h(w) + (1-w) \log d + \lm f(w,d),\label{eq:chMdEl}\\
	\text{where}\quad f(w,d) &=  \eta \! \left(\frac{1-w}{d^2} + w\right) + (d^2 - 1) \; \eta \! \left(\frac{1-w}{d^2} \right).\label{eq:f}
\end{align}
To show this, we recall that the erasure flag of an erasure channel $\EC_{\lambda,d}$ is orthogonal to the input/output space, which allows us to apply \eqref{eq:entropy-orthogonal} to output states of this channel.
Furthermore, the complementary channel of an erasure channel is again an erasure channel, $\EC_{\lambda,d}^c = \EC_{1-\lambda,d}$.

We first compute the mixed input state $\rho_{aa'}=\tr_{rr'}[\psi]_{aa'rr'}$ on $\HC_a\otimes\HC_{a'}$ from \eqref{eq:pure-input-state}:
\begin{align}
	\rho_{aa'} &= (1-w) [0]_{a} \otimes \frac{1}{d} \one_{a'} + \frac{w}{d} \sum_{i,j=1}^d |i,i-1\rangle\langle j,j-1|_{aa'}.
\end{align}
Acting with the erasure channel $\EC_{\lambda,d}$ on the $a'$-system gives a state
\begin{align}
	\sigma_{ab'} &= (\id_a\otimes \EC_{\lambda,d})(\rho_{aa'})\\
	&= (1-\lambda) \rho_{ab'} + \lambda \rho_a \otimes [e]_{b'},\label{eq:sigma-ab}
\end{align}
where we relabeled $a'\to b'$ for $\rho$, and $[e]_{b'}$ denotes the pure erasure flag.
The action of $\MC_{d+1}$ on the $a$-system of $\rho_{ab'}$ produces a state
\begin{align}
	\tau_{bb'} = (\MC_{d+1}\otimes \id_{b'})(\rho_{ab'}) &= (1-w) \MC_{d+1}([0]_a)\otimes \frac{1}{d}\one_{b'} + \frac{w}{d} \sum_{i,j=1}^d \MC_{d+1}(|i\rangle\langle j|_{a})\otimes |i-1\rangle\langle j-1|_{b'}\\
	&= (1-w) \frac{1}{d} \widehat{\one}_{b} \otimes \frac{1}{d} \one_{b'} + w [d]_{b} \otimes \frac{1}{d} \one_{b'},\label{eq:rho-bb}
\end{align}
where we introduced the notation $\widehat{\one}_{b}$ for the identity operator
    on the $d$-dimensional subspace of $\HC_b$ spanned by $\lbrace
    |i\rangle_b\rbrace_{i=0}^{d-1}$ (recall that $\dim\HC_b = d+1 = \dim\HC_a$)
    and $\one_b$ for the identity operator on the $d$-dimensional subspace of
    $\HC_{b'}$ spanned by $\lbrace |i\rangle_{b'}\rbrace_{i=0}^{d-1}$. 
Taking the partial trace over $b'$ in \eqref{eq:rho-bb} yields the state
\begin{align}
	\tau_b = \MC_{d+1}(\rho_a) = (1-w) \frac{1}{d} \widehat{\one}_b + w [d]_b. \label{eq:rho-b}
\end{align}
We then have $\sigma_{bb'} = (\MC_{d+1}\otimes \id_{b'})(\sigma_{ab'}) = (1-\lambda) \tau_{bb'} + \lambda \tau_{b}\otimes [e]_{b'}$, whose entropy we can compute using \eqref{eq:entropy-orthogonal}:
\begin{align}
	S(\sigma_{bb'}) &= h(\lambda) + (1-\lambda) S(\tau_{bb'}) + \lambda S(\tau_b)\\
	&= h(\lambda) + (1-\lambda) [h(w) + (1-w)2\log d + w\log d] + \lambda [h(w) + (1-w)\log d]\\
	&= h(\lambda) + h(w) + (2-\lambda - w)\log d.\label{eq:S-sigma-bb}
\end{align}

For the complementary channel $\MC_{d+1}^c\otimes \EC_{\lambda,d}^c$ we follow a similar strategy, computing first the action of the erasure channel on the $a'$-system of the input state $\rho$:
\begin{align}
	\sigma_{ac'} = (\id_a\otimes \EC_{\lambda,d}^c)(\rho_{aa'}) = \lambda \rho_{ac'} + (1-\lambda) \rho_a \otimes [e]_{c'}.
\end{align}
Let now $|\widehat{\phi}\rangle_{cc'}$ be a maximally entangled state between
    $\HC_{c'}$ and the $d$-dimensional subspace of $\HC_c$ spanned by $\lbrace
    |i\rangle_b\rbrace_{i=0}^{d-1}$. The channel $\MC_{d+1}^c$ acting on the
    $a$-system of $\rho_{ac'}$ yields a state
\begin{align}
	\tau_{cc'} = (\MC_{d+1}^c\otimes \id_{c'})(\rho_{ac'}) &= (1-w) \MC_{d+1}^c ([0]_a) \otimes \frac{1}{d} \one_{c'} + \frac{w}{d} \sum_{i,j=1}^d \MC_{d+1}^c(|i\rangle\langle j|_{a})\otimes |i-1\rangle\langle j-1|_{c'}\\
	&= (1-w) \frac{1}{d} \widehat{\one}_c \otimes \frac{1}{d} \one_{c'} + \frac{w}{d} \sum_{i,j=1}^d |i-1,j-1\rangle\langle i-1,j-1|_{cc'}\\
	&= (1-w) \frac{1}{d} \widehat{\one}_c \otimes \frac{1}{d} \one_{c'} + w \left[\widehat{\phi}\right]_{cc'},\label{eq:rho-cc} \\
    &= \frac{1-w}{d^2} P_{cc'} + \big( w + \frac{1-w}{d^2} \big)\left[\widehat{\phi}\right]_{cc'}, 
\end{align}
where $P_{cc'}:= \widehat{\one}_c \otimes \one_{c'} - [\widehat{\phi}]_{cc'}$
is a projector onto a $(d^2 - 1)$-dimensional subspace of $\HC_c \ot \HC_{c'}$, orthogonal to the support of
$[\widehat{\phi}]_{cc'}$. Thus the last equality above essentially
splits $\tau_{cc'}$ into a sum of two density operators with orthogonal
support.
The marginal of $\tau_{cc'}$ is given by $\tau_c = \frac{1}{d}\widehat{\one}_c$, and so the entropy of $\sigma_{cc'} = \lambda \tau_{cc'} + (1-\lambda)\tau_{c}\otimes [e]_{c'}$ equals
\begin{align}
	S(\sigma_{cc'}) &= h(\lambda) + \lambda S(\tau_{cc'}) + (1-\lambda)S(\tau_c)\\
	& =h(\lambda) + \lambda \left[-\eta\!\left(\frac{1-w}{d^2} + w\right)-(d^2-1)\,\eta\!\left(\frac{1-w}{d^2}\right)\right] + (1-\lambda)\log d\\
	&= h(\lambda) -\lambda f(w,d) + (1-\lambda) \log d,\label{eq:S-sigma-cc}
\end{align}
with $f(w,d)$ as defined in \eqref{eq:f}.

Using \eqref{eq:S-sigma-bb} and \eqref{eq:S-sigma-cc}, we finally arrive at
\begin{align}
	\Dl(\MC_{d+1} \ot \EC_{\lm,d}, \rho_{aa'}) &= S(\sigma_{bb'}) - S(\sigma_{cc'}) = 	h(w) + (1-w) \log d + \lm f(w,d),
\end{align}
which is what we set out to prove.%
In the sequel, we let $\Dl^{*}(\MC_{d+1} \ot \EC_{\lm,d})$ denote the value of $\Dl(\MC_{d+1} \ot
\EC_{\lm,d}, \rho_{aa'})$ maximized over $w\in[0,1]$.

\subsubsection{(Super)-amplification of coherent information}
\label{sec:mdcohinfo}

Similar to our study in Section \ref{sec:superNs}, we focus on the quantity
\begin{align}
	\delta(\MC_{d+1}, \EC_{\lm,d}) =
	\Dl^{*}(\MC_{d+1} \ot \EC_{\lm,d}) - \QC^{(1)}(\MC_{d+1}) -
	\QC^{(1)}(\EC_{\lm,d}) \,. 
\end{align} 
Since $\Dl^{*}(\MC_{d+1}, \EC_{\lm,d}) \leq \QC^{(1)}(\MC_{d+1} \ot \EC_{\lm,d})$,
the positivity of $\delta(\MC_{d+1}, \EC_{\lm,d})$ implies the
(super)-amplification of the coherent information of
$\MC_{d+1} \ot \EC_{\lm,d}$.

For each $d \geq 2$, we perform the optimization for
$\Dl^{*}(\MC_{d+1} \ot \EC_{\lm,d})$ and compute
$\QC^{(1)}(\MC_{d+1})$ numerically to determine if
$\delta(\MC_{d+1}, \EC_{\lm,d})$ is positive.
We find that $\delta(\MC_{d+1}, \EC_{\lm,d})$ starts to turn positive
at some $\lm_{\min}(d) \ll 1/2$, increases to a maximum at $\lm=1/2$,
and then decreases and vanishes as $\lm$ increases to some
$\lm_{\max}(d)$.  These boundaries $\lm_{\min}(d)$ and $\lm_{\max}(d)$ are plotted as functions of $d$ in Fig.~\ref{fig:Q1Q}, while the function $\delta(\MC_{d+1}, \EC_{\lm,d})$ is plotted for some typical values of $d$ in Fig.~\ref{fig:MdSuperEl}.
In addition, $\delta(\MC_{d+1}, \EC_{\lm,d})$ can be proved to be positive for any $0 < \lm < 1$ provided $d$ is sufficiently large, see next subsection. 

Subject to the spin alignment conjecture, $\QC(\MC_{d+1}) = \QC^{(1)}(\MC_{d+1})$. 
From~\eqref{eq:ErQ1Qd},  $\QC^{(1)}(\EC_{\lm,d}) = \QC(\EC_{\lm,d}) = 0$.
Hence, the above (super)-amplification of the coherent information can be
lifted to that of quantum capacity:
\begin{align}
	\QC(\MC_{d+1} \ot \EC_{\lm,d}) \geq \QC^{(1)}(\MC_{d+1} \ot \EC_{\lm,d}) 
	>\QC(\MC_{d+1}) + \QC(\EC_{\lm,d}) \,. 
	\label{eq:QSuperAdd}
\end{align}

\subsubsection{Unconditional amplification of quantum capacity}

We now present a direct proof of amplification and super-amplication of quantum
capacity for $\MC_{d+1}$ and $\EC_{\lm,d}$ without conditioning on the SAC.
The sum of the separate capacities of $\MC_{d+1}$ and $\EC_{\lm,d}$
can be bounded from above using \eqref{eq:Md-Q-bound} and \eqref{eq:ErQ1Qd}:
\begin{align}
	\QC(\MC_{d+1}) + \QC(\EC_{\lm,d}) \leq \log\left(1+\frac{1}{\sqrt{d}}\right) + \max\lbrace (1-2\lambda)\log d,0\rbrace \eqqcolon u(\lambda,d) \,.
	\label{eq:Md-erasure-Q-ub}
\end{align}
We define a quantity
\begin{align}
	\mu(w, d, \lm) := \Dl(\MC_{d+1} \ot \EC_{\lm, d}, \rho_{aa'}) - u(\lambda,d) 
	\label{eq:deltaVal}
\end{align}
so that when $\mu(w, d, \lm)$ is positive, 
we obtain the following chain of inequalities, 
\begin{align}
	\QC(\MC_{d+1}) + \QC(\EC_{\lm,d}) \leq u(\lambda,d) < \Dl^{*}(\MC_{d+1} \ot \EC_{\lm,d}) \leq 
	\QC(\MC_{d+1} \ot \EC_{\lm,d}), 
	\label{eq:Md-amplification}
\end{align}
thus establishing amplification or super-amplification of the quantum capacity
of $\MC_{d+1}$ and $\EC_{\lm,d}$.

Our numerical optimization shows that
$\mu(w,d,\lm) > 0$ 
for $\lambda \in [0.37,0.57]$ and $d=d(\lambda)$ chosen
sufficiently large.
For $\lambda \in [0.37,0.5)$ (and suitably chosen $d$) both channels
$\MC_{d+1}$ and $\EC_{\lm,d}$ have strictly positive quantum capacity, and
hence we find super-amplification of quantum capacity in this parameter regime.
For example, this occurs for $0.4476<\lambda<0.5210$ and $d\geq 8$.
The smallest dimension for which amplification of quantum capacity can be certified is $d=4$ at $\lambda = 1/2$.
This can be seen in Fig.~\ref{fig:Q1Q}, in which we numerically find $\lm^{\QC}_{\min}(d), \lm^{\QC}_{\max}(d)$ for each $d$ so that $\max_w \mu(w,d,\lm) > 0 $ for $\lm^{\QC}_{\min}(d) \leq \lm \leq \lm^{\QC}_{\max}(d)$.
In addition, we find that $\max_w \mu(w,d,\lm)$ depends on $\lm$
similarly as for $\delta(\MC_{d+1}, \EC_{\lm,d})$ (see
Fig.~\ref{fig:MdSuperEl}).  

Finally, we present an analytic proof of amplification and
superamplification of the quantum capacity of
$\MC_{d+1} \ot \EC_{\lm,d}$ for the entire range of erasure probability
$0 < \lm < 1$ using a $\log$-singularity-like argument~\cite{Siddhu21}.
We will show that $\mu(w,d,\lm) > 0$ for some $w$ when $d$ is sufficiently large.  
We consider the three cases $0 < \lm < 1/2$, $\lm=1/2$, and $1/2 < \lm \leq 1$
separately.  
For the first two cases, when $0 < \lm \leq 1/2$, the quantity $\mu(w, d,\lm)$ takes the form
\begin{align}
	\mu(w, d, \lm) \simeq c_0 + c_1 \log d + O\left(\frac{1}{\sqrt{d}}\right),
	\label{eq:dl}
\end{align}
where $c_0 = (1 - \lm) h(w)$ and $c_1 = -w (1- 2 \lm)$.
When $0 < \lm < 1/2$, since $0 < w < 1$, both $c_0 > 0$ and $c_1 < 0$.
Furthermore, $c_0>0$ has a $w$-$\log$-singularity of rate $1 - \lm$ 
while $c_1$ does not, so, $\mu(w, d, \lm) >0$ for sufficient small $w$.
For example, this is achieved by the choice 
\begin{align}
   w = \exp \left( 1 -2 \; \frac{|1{-}2\lm|}{1-\lm} \; \log d \right) \,,
	\label{wStar}
\end{align}
with $\exp$ taken to base $2$.
For $\lm = 1/2$ we have $c_1 = 0$, and letting $w = 1/2$ gives $\mu(w, d, \lm) \simeq 1/2 >0$.

When $1/2 < \lm < 1$,
\begin{align}
	\mu(1-w, d, \lm) \simeq c_0 - c_1 \log_2 d + O\left(\frac{1}{\sqrt{d}}\right).
	\label{eq:dl2}
\end{align}
As argued before, we then have $\mu(1-w, d, \lm)>0$ for small enough $w$;
the choice in~\eqref{wStar} once again works here.
Note that this analysis also shows (super)-amplification 
for coherent information for the same range of $\lm$ mentioned in 
Section \ref{sec:mdcohinfo}.

We conclude with the following observation: For $\lambda = 1/2$ and large $d$,
the sum of the capacities of the two channels becomes arbitrarily small, since
$\QC(\EC_{1/2,d}) = 0$ due to \eqref{eq:ErQ1Qd}, and $\QC(\MC_{d+1})$ has the
vanishing upper bound \eqref{eq:Md-Q-bound}.  Meanwhile, the coherent
information $\Dl(\MC_{d+1} \ot \EC_{1/2,d}, \rho_{aa'})$ tends to $h(w)/2$ as
$d \to \infty$.  At $w=1/2$, this lower bound is exactly equal to $1/2$.
Thus, the channels used jointly retain positive coherent information and
quantum capacity at least $1/2$.
This phenomenon can be considered as a form of ``near-super-activation.''

\begin{figure}[ht]
	\centering
	\begin{tikzpicture}
		\begin{axis}[
			xlabel=$d$,
			ylabel=$\lambda$,
			ylabel style={rotate=-90},
			scale=1.5,
			every axis plot/.append style={line width=1.5pt},
			xmode = log,
			xmax = 2048,
			legend style={at = {(0.97,0.75)},anchor = north east,font=\large},
			grid = both,
			]
			\addplot[name path = upper,mark=none,color=plotmagenta] table[col sep = comma,x=dExp,y=lmMaxQ1] {MdPlusOneErLmNonAddQ1Log.csv}; %
			\addplot[name path = upper,mark=none,color=plotmagenta,dashdotted] table[col sep = comma,x=dExp,y=lmMaxQExp] {MdPlusOneErLmNonAddQLog.csv};
			\addplot[name path = lower,mark=none,color=plotblue,dashdotted] table[col sep = comma,x=dExp,y=lmMinQExp] {MdPlusOneErLmNonAddQLog.csv};
			\addplot[name path = lower,mark=none,color=plotblue] table[col sep = comma,x=dExp,y=lmMinQ1] {MdPlusOneErLmNonAddQ1Log.csv};%
			\legend{$\lambda_{\mathrm{max}}(d)$,$\lambda_{\mathrm{max}}^{\QC}(d)$,$\lambda_{\mathrm{min}}^{\QC}(d)$,$\lambda_{\mathrm{min}}(d)$};
		\end{axis}
	\end{tikzpicture}
	\caption{
		Plot of the region of super-additivity of coherent information and quantum capacity of the quantum channel $\MC_{d+1} \ot \EC_{\lambda,d}$.
		The solid lines are the minimal values $\lambda_{\mathrm{min}}(d)$ (blue) and maximal values $\lambda_{\mathrm{max}}(d)$ (magenta) between which $\delta(\MC_{d+1}, \EC_{\lambda,d}) > 0$, that is, $\MC_{d+1} \ot \EC_{\lambda,d}$ has super-additive coherent information.
		The dashed lines are the minimal values $\lambda^{\QC}_{\mathrm{min}}(d)$ (blue) and maximal values $\lambda^{\QC}_{\mathrm{max}}(d)$ (magenta) between which $\mu(w,d,\lm)>0$, that is, $\MC_{d+1} \ot \EC_{\lambda,d}$ has super-additive quantum capacity.
		This figure is identical to Fig.~2 in the main text and reproduced here for convenience.
	}
	\label{fig:Q1Q}
\end{figure}
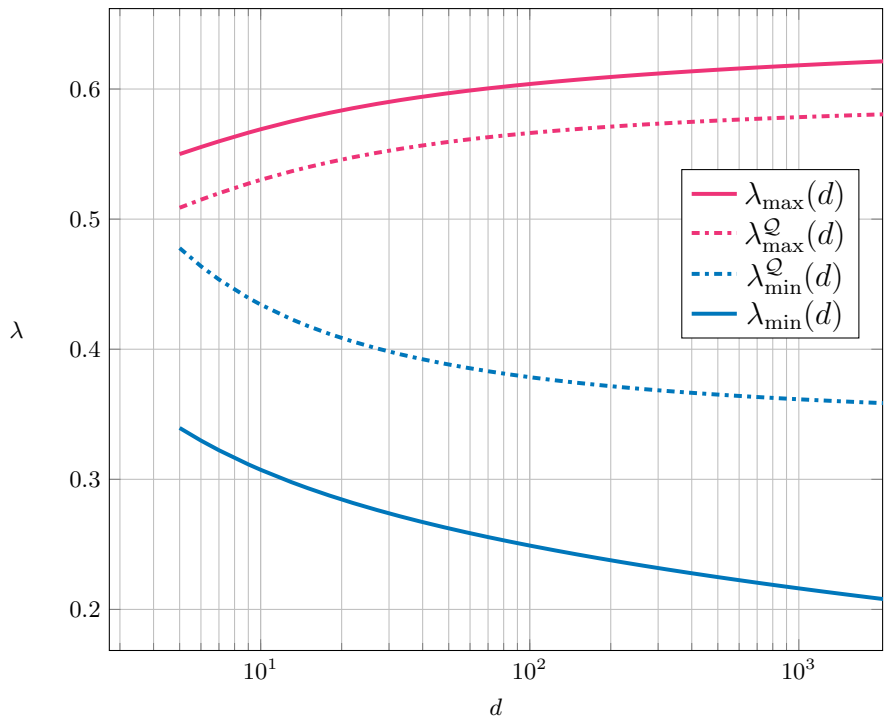

\begin{figure}[ht]
	\centering
	\begin{tikzpicture}
		\begin{axis}[
			title = {$d=10$},
			xlabel=$\lambda$,
			every axis plot/.append style={line width=1.5pt},
			legend style={at = {(0.02,0.02)},anchor = south west,font=\small},
			legend cell align={left},
			grid = both,
			xmin = 0,
			xmax = 1,
			ymin = -0.7,
			]
			\addplot[color = plotgray,semithick] {0};
			\addplot[name path = lower,mark=none,color=plotgreen] table[col sep = comma,x=lm,y=diffQ1] {d10NonAddData.csv};
			\addplot[name path = upper,mark=none,color=plotorange] table[col sep = comma,x=lm,y=diffQ] {d10NonAddData.csv};
			\legend{,$\delta(\MC_{d+1}{,}\EC_{\lambda{,}d})$,$\Dl^*(\MC_{d+1} \ot \EC_{\lm{,}d}) - u(\lm{,}d)$};
		\end{axis}
	\end{tikzpicture}\hfill
	\begin{tikzpicture}
		\begin{axis}[
			title = {$d=50$},
			xlabel=$\lambda$,
			every axis plot/.append style={line width=1.5pt},
			legend style={at = {(0.98,0.75)},anchor = north east},
			grid = both,
			xmin = 0,
			xmax = 1
			]
			\addplot[color = plotgray,semithick] {0};
			\addplot[name path = lower,mark=none,color=plotgreen] table[col sep = comma,x=lm,y=diffQ1] {d50NonAddData.csv};
			\addplot[name path = upper,mark=none,color=plotorange] table[col sep = comma,x=lm,y=diffQ] {d50NonAddData.csv};
		\end{axis}
	\end{tikzpicture}
	\caption{
		Plotting $\delta(\MC_{d+1}, \EC_{\lm,d})$ (green) and
		$\Dl^*(\MC_{d+1} \ot \EC_{\lm,d}) - u(\lm,d)$ (orange) for 
		$d=10$ (left) and $d=50$ (right).  When these quantities are positive,
		$\MC_{d+1} \ot \EC_{\lm,d}$ has (super)-amplification of coherent
		information and quantum capacity respectively.}
	\label{fig:MdSuperEl}
\end{figure}
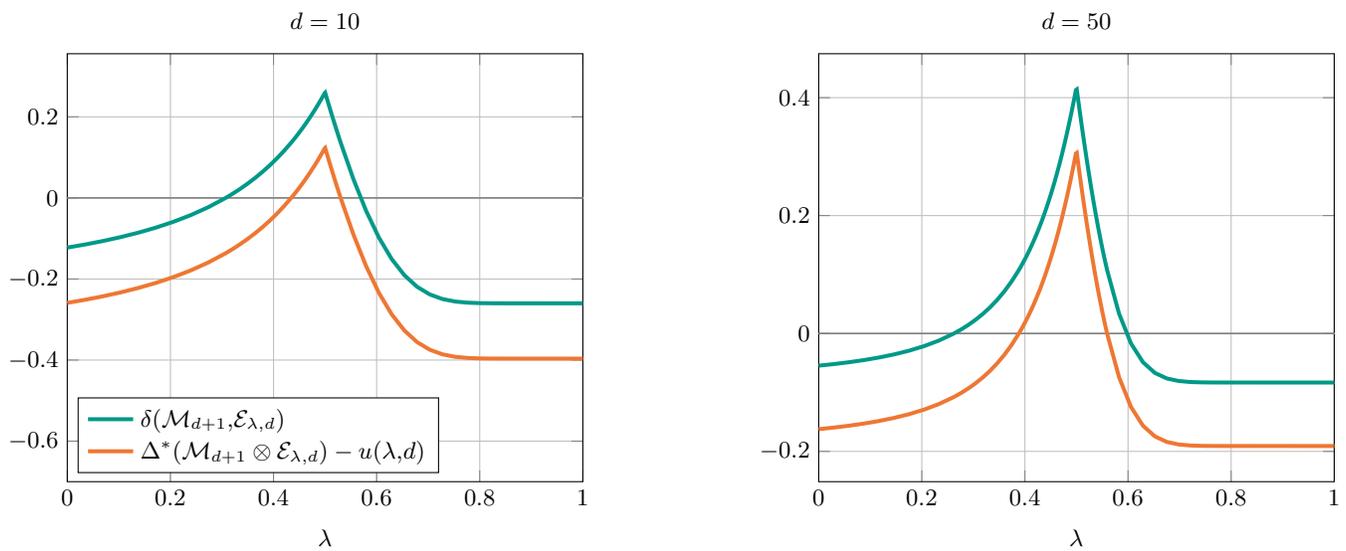

\end{document}